\documentclass[fleqn,usenatbib]{mnras}

\usepackage{newtxtext,newtxmath}

\usepackage[T1]{fontenc}
\usepackage{ae,aecompl}

\usepackage{graphicx}	
\usepackage{amsmath}	
\usepackage{amssymb}	

\newcommand{\cmmm}{\text{cm}^{-3}}

\newcommand{\msunyr}{\text{M}_\odot\,\text{yr}^{-1}}

\newcommand{\HH}{\text{H}_2}          
\newcommand{\HM}{{\rm H}^{-}}     
\newcommand{\HP}{{\rm H}^{+}}     
\newcommand{\HII}{H{\sc ii}~}    
\newcommand{\e}{{\rm e}^{-}}     

\newcommand{\nh}{n_{\rm H}}

\newcommand{\msun}{\text{M}_\odot}

\newcommand{\tff}{t_{\rm ff}}

\newcommand{\rvir}{R_{\rm vir}}
\newcommand{\Tvir}{T_{\rm vir}}
\newcommand{\mvir}{M_{\rm vir}}

\newcommand{\mbe}{M_{\rm BE}}
\newcommand{\menc}{M_{\rm enc}}

\newcommand{\cs}{c_{\rm s}}

\newcommand{\vrad}{v_{\rm rad}}

\newcommand{\mstardot}{\dot{M}_\star}
\newcommand{\mstar}{M_\star}
\newcommand{\rstar}{R_\star}

\newcommand{\mh}{m_{\rm H}}

\newcommand{\Racc}{R_{\rm acc}}

\newcommand{\nth}{n_{\rm th}}
\newcommand{\mdot}{\dot{M}}

\newcommand{\taug}{{\boldsymbol \tau}_{\rm grav}}
\newcommand{\taup}{{\boldsymbol \tau}_{\rm pres}}
\newcommand{\tgrav}{t_{\rm grav}}
\newcommand{\tpres}{t_{\rm pres}}
\newcommand{\rIF}{r_{\rm IF}}
\newcommand{\Nion}{\dot{N}_{\rm ion}}

\title[Assembly of SMBH seeds]{Assembly of supermassive black hole seeds}

\author[Becerra et al.]{\parbox{17.5cm}{Fernando Becerra$^1$\thanks{E-mail: fbecerra@cfa.harvard.edu}, Federico Marinacci$^{1,2}$, Volker Bromm$^3$ and Lars E. Hernquist$^1$}
\\
\\$^1$ Harvard-Smithsonian Center for Astrophysics, 60 Garden Street, Cambridge, MA 02138, USA
\\$^2$ Kavli Institute for Astrophysics and Space Research, Massachusetts Institute of Technology, Cambridge, MA 02139, USA
\\$^3$ Department of Astronomy, The University of Texas at Austin, TX 78712, USA}

\pubyear{2018}

\begin{document}
\label{firstpage}
\pagerange{\pageref{firstpage}--\pageref{lastpage}}
\maketitle

\begin{abstract}
We present a suite of six fully cosmological, three-dimensional simulations of the collapse of an atomic cooling halo in the early Universe. We use the moving-mesh code {\sc arepo} with an improved primordial chemistry network to evolve the hydrodynamical and chemical equations. The addition of a strong Lyman-Werner background suppresses molecular hydrogen cooling and permits the gas to evolve nearly isothermally at a temperature of about 8000~K. Strong gravitational torques effectively remove angular momentum and lead to the central collapse of gas, forming a supermassive protostar at the center of the halo. We model the protostar using two methods: sink particles that grow through mergers with other sink particles, and a stiff equation of state that leads to the formation of an adiabatic core. We impose threshold densities of $10^8$, $10^{10}$, and $10^{12}\,\cmmm$ for the sink particle formation and the onset of the stiff equation of state to study the late, intermediate, and early stages in the evolution of the protostar, respectively. We follow its growth from masses $\simeq 10\,\msun$ to $\simeq 10^5\,\msun$, with an average accretion rate of $\langle\mstardot\rangle \simeq 2\,\msunyr$ for sink particles, and $\simeq 0.8 - 1.4\,\msunyr$ for the adiabatic cores. At the end of the simulations, the \HII region generated by radiation from the central object has long detached from the protostellar photosphere, but the ionizing radiation remains trapped in the inner host halo, and has thus not yet escaped into the intergalactic medium. Fully coupled, radiation-hydrodynamics simulations hold the key for further progress.
\end{abstract}




\section{Introduction}
\label{sec:intro}

Frontier observations of quasars at redshift z $\gtrsim 6$ suggest the existence of supermassive black holes (SMBHs) with masses $\gtrsim 10^9\,\msun$, when the Universe was less than one billion years old \citep{Fan_2003, Fan_2006, Mortlock_2011, Wu_2015, Banados_2018}. These SMBHs most likely grew from smaller seed BHs, although the origin of these seeds is still unclear \citep{Haiman_2006, Haiman_2009, Bromm_Yoshida_2011, Greene_2012, Volonteri_2012, Volonteri_Bellovary_2012, Greif_2015, Johnson_2016, Latif_2016b, Smith_2017b}. The two most promising pathways for the formation of BH seeds at high redshift are the remnants of massive Population III stars \citep{Madau_2001, Li_2007, Johnson_2012, Alexander_2014}, and the direct collapse of primordial gas in haloes with virial temperatures $T_{\rm vir}\gtrsim 10^4\,$K, the so-called atomic cooling haloes \citep{Bromm_2003, Begelman_2006, Spaans_2006}. A third formation scenario, invoking the high-velocity mergers of massive proto-galaxies, does not require the absence of metals or other low-temperature coolants \citep[e.g.][]{Mayer_2010, Inayoshi_2015}.

In the direct collapse scenario, gas falls into the center of haloes where cooling by molecular hydrogen and metal lines to below $\simeq 10^4$ K has been suppressed. The destruction of molecular hydrogen can be achieved by its photo-dissociation due to external soft ultraviolet (UV) background radiation in the Lyman-Werner (LW) bands \citep{Omukai_2001, Bromm_2003, Volonteri_2005, Spaans_2006, Schleicher_2010, Johnson_2013, Agarwal_2016, Habouzit_2016, Johnson_2017}. In such a case, the dominant coolant is atomic hydrogen \citep{Omukai_2001, Oh_2002}, and the gas follows a nearly isothermal collapse at temperatures around the virial value $\Tvir \simeq 10^4$ K: first, due to Lyman-$\alpha$ cooling up to densities $\nh \simeq 10^6\,\cmmm$, where the gas becomes optically thick to Ly$\alpha$ radiation, and then due to $\HM$ bound-free and free-free emission \citep{Regan_2009, Latif_2013a, Inayoshi_2014, Becerra_2015, Becerra_2017, Chon_2016}. As the gas keeps contracting, it becomes optically thick to $\HM$ continuum emission around $\nh \simeq 10^{17}\,\cmmm$, at which point the gas evolves adiabatically and forms a massive protostar at the center of the halo \citep{Inayoshi_2014, VanBorm_2014, Becerra_2015, Latif_2016a}. Since the accretion rate in a Jeans-unstable cloud scales as $\mdot \propto T^{3/2}$ and the gas in an atomic cooling halo can reach temperatures $T \simeq 10^4$ K, very high values for the accretion rate ($\simeq 1\,\msunyr$) are achieved. As a consequence, the protostar can grow to $\simeq 10^5 - 10^6\,\msun$ and become a supermassive star in about a million years \citep{Regan_2009, Latif_2013b}.

Previous high-resolution studies have been able to describe in detail the initial assembly of the protostar up to masses of $\mstar \simeq 0.1\,\msun$, but, due to the high densities involved in this process, they only follow its evolution for a few years \citep{Inayoshi_2014, VanBorm_2014, Becerra_2015, Latif_2016a}. In order to investigate the growth of such protostar for longer times, different techniques have been adopted to avoid the creation of high-density regions that prohibitively slow down the simulations. For example, \citet{Regan_2009, Latif_2013b, Choi_2015} used a pressure floor beyond a certain resolution level, which limited the maximum density to $\nh \simeq 10^6 - 10^{12}\,\cmmm$. Similarly, \citet{Latif_2013d, Shlosman_2016, Regan_2018} employed sink particles that replace gas above a maximum density of $\nh \simeq 10^5 - 10^8\,\cmmm$. One disadvantage of these methods is the limited resolution in regions where density is highest, hence they only describe well the processes on large scales.

In this work, we reconstruct the assembly process of SMBH seeds at high redshifts from masses $\mstar \simeq 10\,\msun$ to $\mstar \simeq 10^4-10^5\,\msun$. For that purpose, we perform a series of simulations starting from cosmological initial conditions, employing both sink particles and an artificially-stiffened equation of state to model the central object. Additionally, we impose different maximum densities $\nh = 10^8$, $10^{10}$, and $10^{12}\,\cmmm$ to study the late, intermediate, and early stages of the formation of seed BHs, respectively. With this approach, we describe the full picture of the assembly of SMBH seeds. These objects will be prime targets of next-generation observational facilities \citep{Pacucci_2015, Dayal_2017, Natarajan_2017}, such as the {\em James Webb Space Telescope (JWST)}, the ATHENA X-ray mission \citep[e.g.][]{Valiante_2018}, and the Laser Interferometer Space Antenna (LISA) gravitational-wave observatory \citep[e.g.][]{Sesana_2011}.

Our paper is organized as follows. In Section \ref{sec:sims}, we describe the simulation setup, the chemistry and cooling network, and the techniques used to model the central object. In Section \ref{sec:results}, we analyze the simulation and discuss the collapse of the gas cloud at the center of the halo, the formation and evolution of the central object, the redistribution of angular momentum in the surrounding gas, and the moment when radiation breakout occurs. We discuss the caveats of this work in Section \ref{sec:caveats} and, finally, in Section \ref{sec:summary} we summarize and draw conclusions.

\section{Numerical Methodology}
\label{sec:sims}

We investigate the collapse of gas in an atomic cooling halo exposed to a UV background radiation by performing three-dimensional, cosmological hydrodynamical simulations. Specifically, we use the moving-mesh code {\sc arepo} \citep{Springel_2010} with the primordial chemistry network described in detail in \citet{Greif_2014}. In the following, we briefly summarize the procedure used in \citet{Becerra_2015}, and describe the modifications added in this work.

\subsection{Simulation set-up}
\label{subsec:methodology}

We initialize a dark matter (DM)-only simulation at redshift $z=99$ in a $\Lambda$ cold dark matter ($\Lambda$CDM) cosmology. We place 512$^3$ DM particles of mass $\simeq 2.2\times10^3\,\msun$ in a 2 Mpc (comoving) box with a softening length of $\simeq 195$ pc (comoving). We stop the simulation once the first halo to reach a virial mass $\simeq 10^8\,\msun$ collapses. This occurs at $z_{\rm coll} \simeq 11.9$, at which point the halo has a virial mass of $\mvir \simeq 1.5 \times 10^8\,\msun$, a virial radius of $\rvir \simeq 1.4$ kpc, a virial temperature of $\Tvir \simeq 3.5\times10^{4}$ K, and a spin parameter $\lambda \simeq 0.04$.

Once the first $10^8\,\msun$ halo has collapsed, we identify the particles belonging to such halo and trace them back to their initial conditions.  We then increase the resolution of the region enclosing the halo by replacing each DM particle by 64 less-massive DM particles and 64 mesh-generating points. With this setup we are able to achieve a DM particle mass of $M_{\rm dm,ref} \simeq 28~\msun$, and a refined cell mass given by $M_{\rm gas,ref} \simeq 6\,\msun$. In addition, we include a strong radiation background that photo-dissociates molecular hydrogen, resulting in a nearly-isothermal evolution of the collapsing gas (for more details see Section \ref{subsec:chemistry}). We then stop the simulation once the first cell exceeds a density of $\nh \simeq 10^8\,\cmmm$, and proceed with a zoom-in simulation by extracting the central 20~pc of the original box. Assuming a temperature of $T \simeq 10^4\,{\rm K}$, this box size ensures that material on the edge of the box will not have enough time to fall to the centre of the cut-out, $t \simeq 20\,{\rm pc} / \cs \simeq 2{\rm Myr}$, where $\cs$ is the sound speed. The artificial boundary will thus, for the duration of the simulation, not affect the halo center, where the protostellar object is created and evolved according to the methodology described in Section \ref{subsec:object_modeling}.

To achieve such high densities, we enforce the Truelove criterion \citep{truelove_1997} for refinement, which indicates that the Jeans length needs to be resolved by at least four cells to capture gravitational instabilities. In addition, previous studies have found that the Jeans length must be resolved by at least 32 cells in order to adequately describe the effects of turbulence \citep{Federrath_2011, Safranek_2012, Turk_2012, Latif_2013a}. Consequently, to fulfill both requirements, and to be on the safe side, we employ a higher resolution of 64 cells per Jeans length, which is evaluated by using $T_{\rm min} = 5000\,$K for cells with $T \le T_{\rm min}$, but the actual temperature for cells with $T>T_{\rm min}$. Following this strategy, the maximum spatial resolution reached is $\simeq 10^{-5}\,$pc ($\simeq 2\,{\rm au}$) for our highest resolution simulation.

\begin{figure*}
\begin{center}
\includegraphics[scale=1.1]{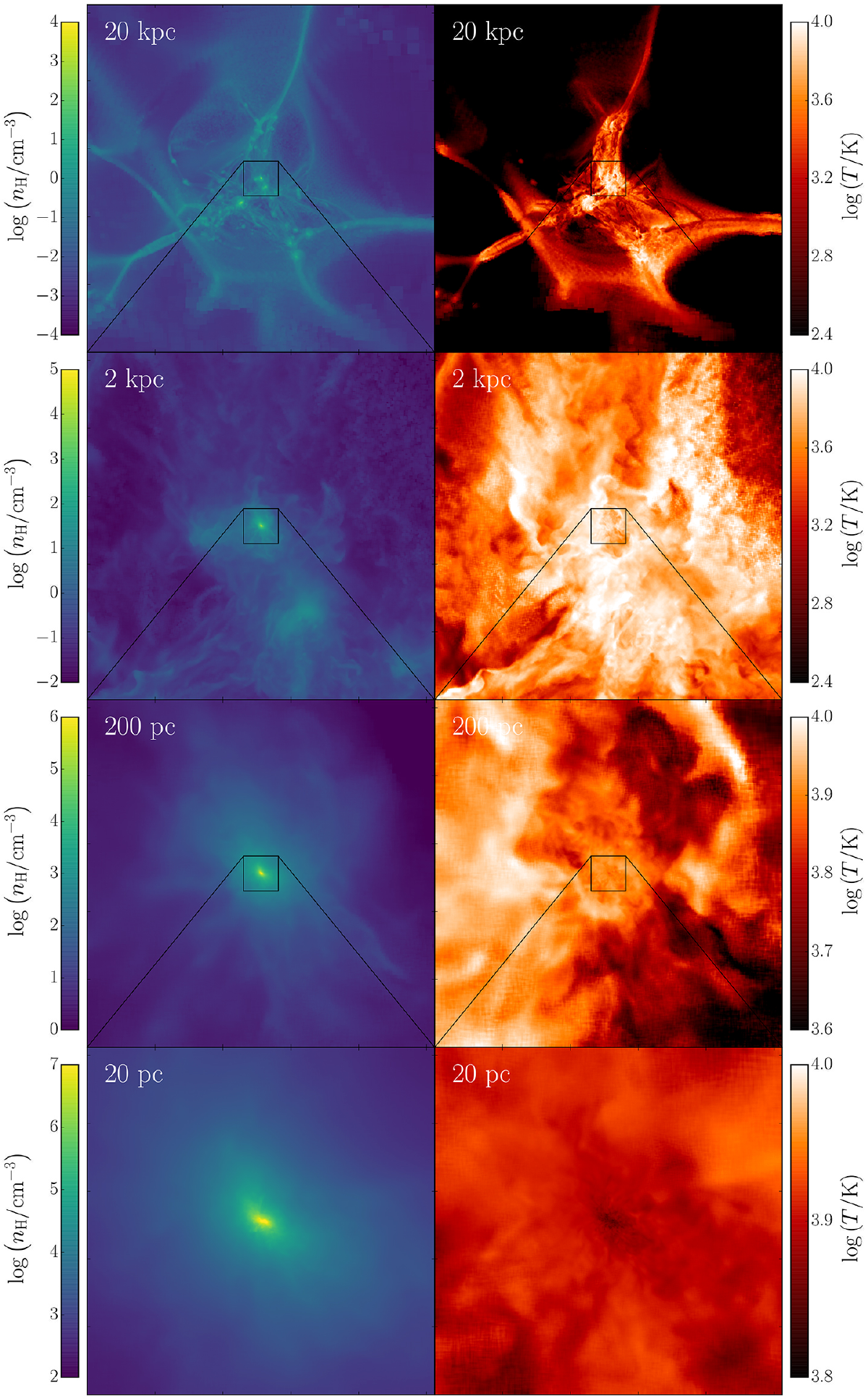}
\caption{Overview of the cosmological simulation once the highest density cell reaches $10^8\,\cmmm$. We show density (left) and temperature (right) projections weighted with the square of the density along the line of sight for box sizes of 20 kpc, 2 kpc, 200 pc, and 20 pc, from top to bottom. At large scales, the central halo is surrounded by smaller haloes and filaments, which are characteristic of the cosmic web. Closer to the center, the structure of the cloud changes from highly irregular due to turbulence to nearly spherical (bottom panel).}
\label{fig:collapse}
\end{center}
\end{figure*}

\subsection{Chemistry}
\label{subsec:chemistry}

The chemical and thermal model is based on the implementation of \citet{Greif_2014} and \citet{Becerra_2015}. Here, we briefly summarize the main components of the network. We employ a non-equilibrium solver for five species (H, $\HH$, $\HM$, $\HP$, $\e$), and include reactions such as the formation of $\HH$ via associative detachment as well as three-body reactions, destruction of $\HH$ via collisions and photodissociation, and the formation and destruction of $\HP$ by collisional ionizations and recombinations.

As in \citet{Becerra_2015}, the cooling processes included are $\HH$ line cooling, $\HH$ collision-induced emission, Ly-$\alpha$ cooling, and inverse Compton cooling. One limitation of that work was the assumption that the optically thin regime for atomic hydrogen cooling extends up to densities $\simeq 10^{16}\,\cmmm$, while in reality gas becomes optically thick to Ly-$\alpha$ at densities of $\simeq 10^{6}\,\cmmm$. To improve on that, we have added an artificial cutoff to Ly-$\alpha$ radiation at such density, and from then on free-free and bound-free $\HM$ continuum cooling dominate. The details and formulae used in the implementation of $\HM$ cooling are described in the appendix of \citet{Becerra_2017}.

Additionally, we have included a strong Lyman-Werner (LW) background radiation field that dissociates $\HH$ via the Solomon process \citep{Abel_1997}. Previous studies have found that photo-dissociation of molecular hydrogen in the progenitors of an atomic cooling halo requires a LW flux with $J_{\rm 21}\gtrsim 10^3$ \citep{Omukai_2001, Johnson_2007, Dijkstra_2008, Latif_2013b, Wolcott-Green_2011, Agarwal_2016}. Here, we assume a constant LW flux of $J_{\rm 21}=10^5$ for a blackbody spectrum with $T_{\rm rad}=10^5\,{\rm K}$. Due to this radiation, $\HH$ cooling does not become important during the collapse of the halo, and hence its evolution is mainly determined first by atomic hydrogen Ly-$\alpha$ cooling, and then by $\HM$ cooling.

\subsection{Modeling the central object}
\label{subsec:object_modeling}

Modeling the formation and evolution of a massive protostar from beginning to end implies reaching extremely high densities, which makes the simulations computationally expensive. Since our goal is to describe the evolution of such object until it achieves a mass of $\simeq 10^4- 10^6\,\msun$, we need a way to bypass this restriction. To accomplish this, we represent the central object employing two different approaches commonly used in the context of direct collapse BH formation: sink particles \citep[e.g.][]{Latif_2013d, Shlosman_2016, Regan_2018} and an artificially-stiffened equation of state \citep[e.g.][]{Regan_2009, Latif_2013a}. Both involve a tradeoff: we lose resolution at the center of the object in order to follow its evolution for a longer period. Here we describe the specific implementation of both methods.

\subsubsection{Sink particles}
\label{subsubsec:sinks}

We pre-set a threshold density, $\nth$, and create a so-called sink particle \citep[e.g.][]{Bate_1995, Bromm_2002} every time a gas element reaches that value. Once that occurs, we take out all the gas cells within some prescribed radius (centered on the densest cell) that are active at that timestep. The mass of the newly-created sink particle is then the total mass removed, while its position and velocity are determined by the position and velocity of the center of mass of those cells. With those values, and using the temperature of the progenitor cell, we then proceed to calculate an accretion radius, $\Racc$, based on the formulation of \citet{Becerra_2017}, and assign it to the sink particle. This process is repeated and the accretion radius is re-calculated and updated every time the sink particle becomes active. This occurs quite often since we force the sink particle to have a timestep that is the minimum between the one used for the gravity solver and the smallest between gas cells inside the accretion radius. This procedure effectively results in a variable accretion radius as a function of the sink mass. In this way, we ensure that, at the beginning, $\Racc$ is determined by the threshold density, but as the object grows in mass the accretion radius transitions to the Bondi radius.

To avoid problems with the mesh reconstruction algorithm in {\sc arepo}, sinks are not allowed to accrete the surrounding gas cells directly. Instead, their growth in mass occurs through mergers with other sink particles. For that purpose, we check if any sink particles are inside the accretion radius of a neighboring sink particle. If that is the case, one of the sink particles is removed and its mass and momentum are transferred to the other one. As a result, the accretion rate of sink particles is fully determined by mergers with other sinks and we do not impose any value a priori, in contrast to previous studies \citep[e.g.][]{Latif_2013d, Shlosman_2016}. This also implies that we can have multiple sink particles for short moments during the evolution of the system. Some of them could be ejected due to gravitational slingshot interactions between them \citep[e.g.][]{Bate_2003}, while others may be accreted because of viscous forces \citep[e.g.][]{Hosokawa_2016}. Following this approach we reach a resolution of $\mstar \simeq 3\,\msun$ for the initial mass of the sink using $\nth = 10^{12}\,\cmmm$.

Because all gas elements with densities larger than $\nth$ are replaced by sink particles, this approach allows us to avoid using computational resources on the evolution of high density regions. As a result, the timesteps associated with high-density cells does not become prohibitively small, which allows us to evolve the system for a longer period of time after the formation of the central object. In contrast, the removal of gas cells implies that this methodology might not properly resolve the inner boundary conditions and torques around the sink particle, which might influence the redistribution of gas around the object and hence its accretion rate.

More sophisticated implementations of sink particles have been suggested in previous studies. For example, some of them include additional checks such as that the gas is converging, Jeans unstable, and gravitationally bound in order to form a sink particle \citep[e.g.][]{Federrath_2010}. Here, we have chosen to employ just the threshold density as a proxy for the creation of sink particles to be consistent with the definition used in \citet{Becerra_2015}, and to make the comparison with the adiabatic objects fairer (see Section \ref{subsubsec:core}). In the same way, we have not enforced any accretion rate formalism a priori, such as the Bondi-Hoyle prescription, although the accretion radius depends implicitly on the Bondi radius. We have made this choice mainly because we want to model sink particles based on just mass and accretion radius, while other characteristics (such as the accretion rate) will be obtained as a result of those properties.
\begin{figure}
\begin{center}
\includegraphics[scale=0.37]{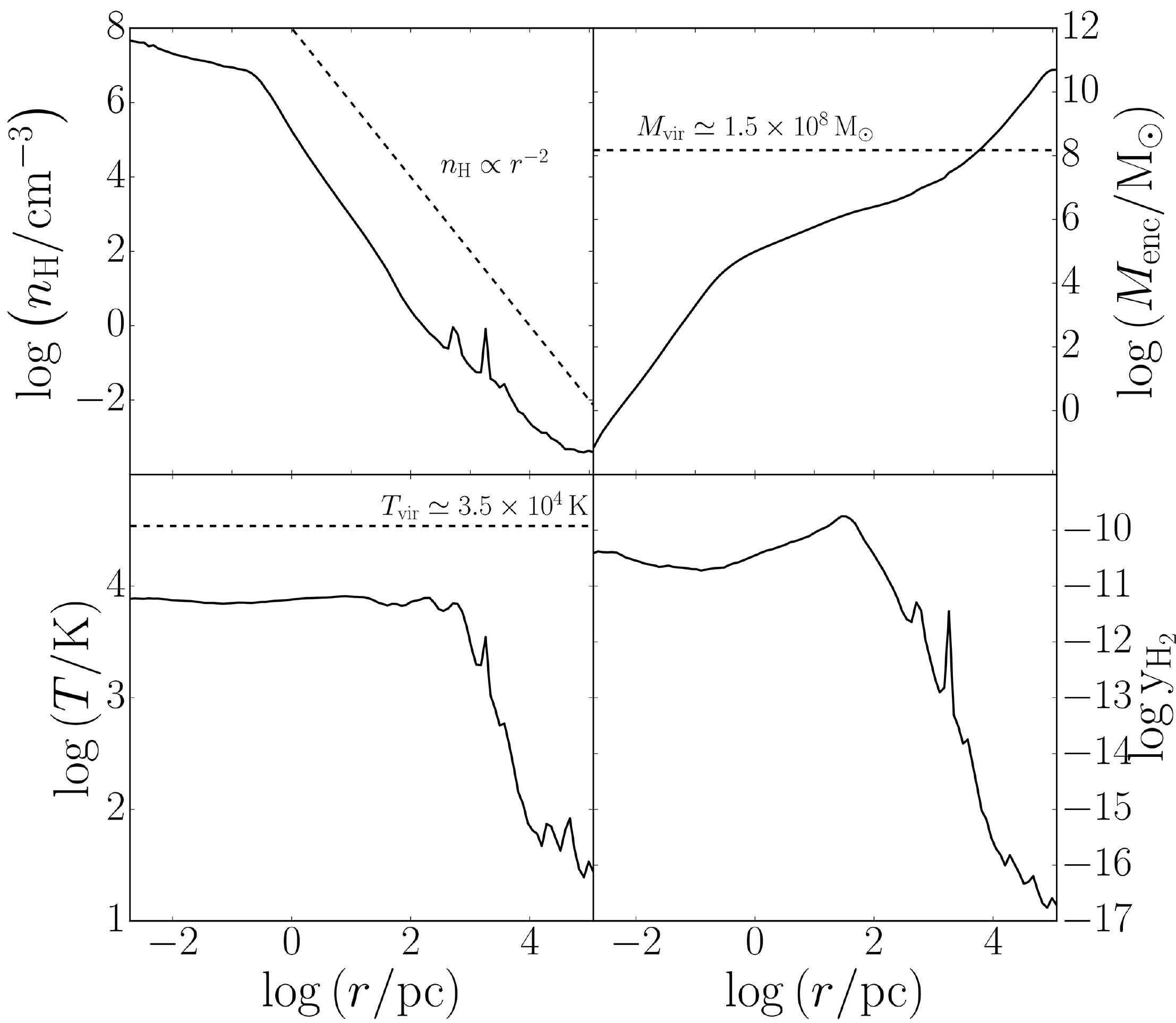}
\caption{Environment around the initial protostar. Radial profiles for the  mass-weighted number density of hydrogen nuclei, enclosed gas mass, temperature, and $\HH$ abundance once the highest density cell reaches $10^8\,\cmmm$. As the halo collapses, its temperature increase to nearly the virial value, while at the same time $y_{\HH}$ reaches a maximum of $\simeq 10^{-10}$. This quantity remains low thanks to the strong LW background radiation, which results in a nearly isothermal evolution of the gas with $T \simeq 8000$\,K, characterized by a density profile $\nh \propto r^{-2}$.}
\label{fig:radial_profiles}
\end{center}
\end{figure}

\begin{figure*}
\begin{center}
\includegraphics[scale=0.95]{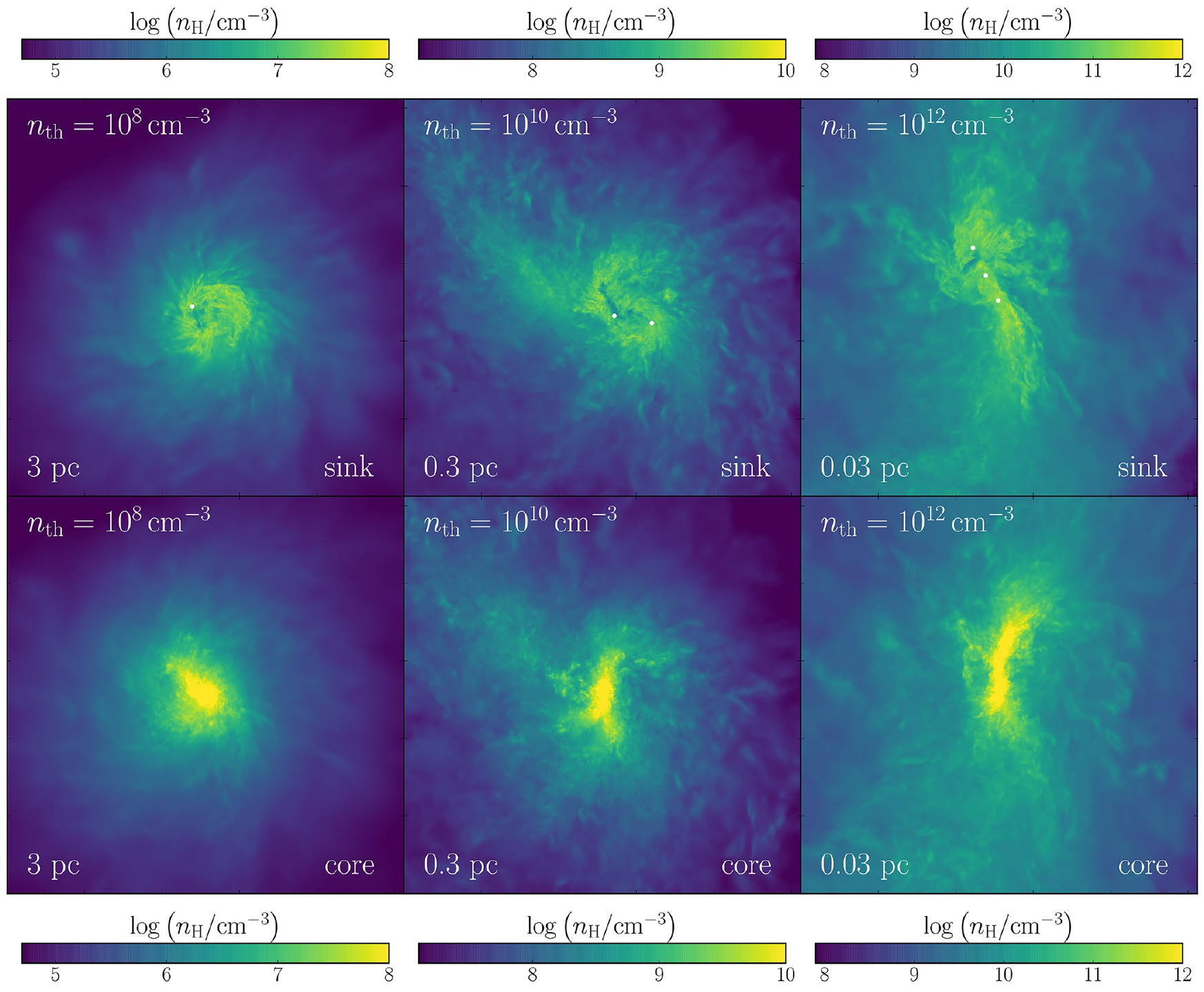}
\caption{Projected number density of hydrogen nuclei at the end of each run for sink (top) and artificial core (bottom) simulations. For the sink particles, their location is represented by white dots. Columns represent threshold densities of $\nth = 10^8$ (left), $10^{10}$ (middle), and $10^{12}\,\cmmm$ (right), with box sizes of 3 pc, 0.3 pc, and 0.03 pc, respectively. Different scales reveal dissimilar morphologies of the gas around the central object. For example, in the left panels, a disk-like structure can be recognized, while at scales $\lesssim 0.01$~pc the gas tends to form more elongated structures. Note that the accretion onto sink particles produces low-density voids, which are not seen in the case of the core simulations.}
\label{fig:nh_snapshots}
\end{center}
\end{figure*}

\subsubsection{Stiff equation of state}
\label{subsubsec:core}

An alternative approach to model the central object is to artificially introduce an exponential cut-off in the cooling rate above a pre-set threshold density, $\nth$. This effectively models the central protostar as an opaque hydrostatic core in which the gas elements with densities higher than $\nth$ evolve adiabatically, thus arresting the dynamical collapse. Here, we follow the implementation described in \citet{Hirano_2017}, where they introduce an artificial opacity, $\tau_{\rm art}$, that depends on the local number density defined by
\begin{equation}
\tau_{\rm art} = \left(\frac{\nh}{\nth}\right)^2,
\end{equation}
and the corresponding escape fraction, $\beta_{\rm esc.art}$, as
\begin{equation}
\beta_{\rm esc.art} = \frac{1-\exp{(-\tau_{\rm art})}}{\tau_{\rm art}}.
\end{equation}
Then, we proceed to multiply all cooling rates by this escape fraction to suppress them above $\nth$ such that $\Lambda = \beta_{\rm esc.art} \times \Lambda_{\rm cool}$. By reducing the cooling rate above $\nth$, gas cells in dense regions experience compressional heating and form an adiabatic core.
To fully characterize the central object, we need to define its radius. To that extent, we follow the approach of \citet{Greif_2012}, in which the photospheric radius of Pop~III stars is determined as the point where the optical depth exceeds unity. Here, $\tau_{\rm art} = 1$ corresponds to the condition where the density reaches the threshold density. Hence, the protostellar radius can be computed using radial profiles and choosing the location where the gas density reaches $\nth$.

On the one hand, this allows us to simplify the hydrodynamics governing the evolution of dense regions, which significantly reduces the computational cost of the simulations. On the other hand, as the central object collapses, cells keep being refined and the highest density increases beyond $\nth$, but at a reduced rate, so that the overall dynamics can be followed for a much longer duration. Eventually, however, the corresponding timestep becomes smaller and smaller, rendering the simulation too expensive to continue.

\begin{figure*}
\begin{center}
\includegraphics[scale=0.95]{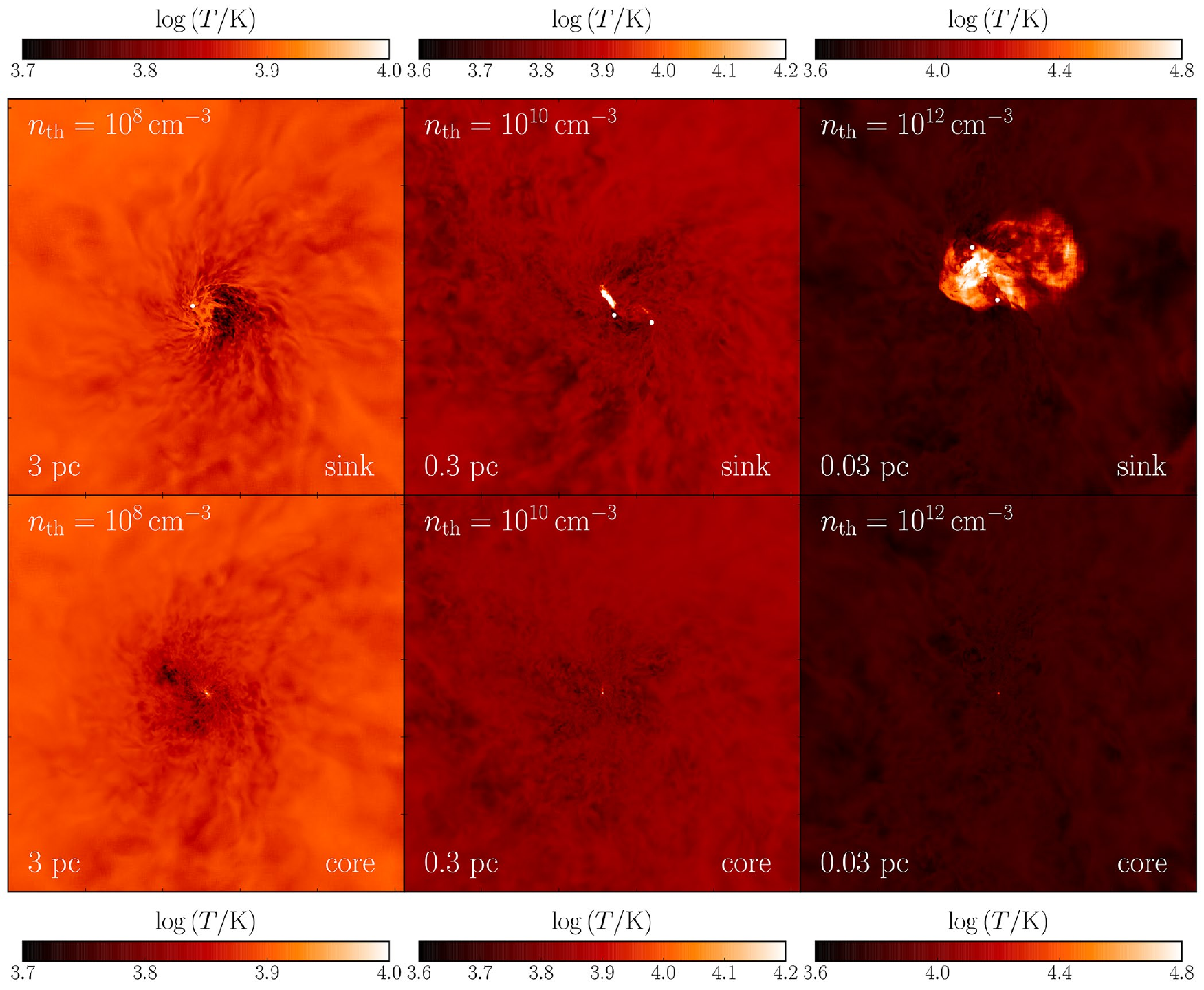}
\caption{Temperature projections at the end of each run for the simulations, threshold densities, and box sizes indicated in Figure \ref{fig:nh_snapshots}. The temperature evolution at these scales is roughly isothermal, although there is an increase at the center. In the case of the sink simulations, this is caused by the dynamical interactions between sink particles that heat up the surrounding gas, while in the case of the core simulations the rise in temperature is due to the adiabatic evolution of the hydrostatic object. Note that in this case, the temperature rise at the center of the image is barely visible on the scale of the projection.}
\label{fig:temp_snapshots}
\end{center}
\end{figure*}

\section{Results}
\label{sec:results}

We present a suite of six simulations using both approaches to model the newly-formed protostar (``the central object''): sink particles and hydrostatic cores. We use threshold densities of $\nth = 10^8$, $10^{10}$, and $10^{12}\,\cmmm$ for each one of them to describe the late, intermediate, and early stages in the evolution of the growing protostar, respectively. By analyzing the behavior of the object at different stages of its life, we can obtain a fuller picture of the buildup of supermassive black hole seeds.

\subsection{Initial collapse}
\label{subsec:collapse}

Figure \ref{fig:collapse} shows the hydrogen number density (left) and temperature (right) projections at the end of the cosmological parent simulations, once the highest density cell reaches $\nh \simeq 10^8\,\cmmm$. From top to bottom, we show box sizes of 20 kpc, 2 kpc, 200 pc, and 20 pc (physical). At large scales we can clearly distinguish the cosmic web surrounding the central halo, composed of filaments and less massive haloes where they intersect. At scales $\gtrsim 2$ kpc turbulence dominates, which causes the cloud morphology to be highly irregular. During the collapse process, the gas temperature inside the virial radius $\rvir \simeq 1.4$~kpc increases to values around $10^4$~K. As we move deeper inside the halo, we see that the morphology changes to a nearly spherical object at $\simeq 20$~pc, and that there is a slight decrease in temperature at scales less than a few pc. The latter is mainly due to the addition of $\HM$ cooling, as noted by \citet{Inayoshi_2014}.

A different approach to study the initial collapse is by plotting different halo properties as a function of radius. In Figure \ref{fig:radial_profiles}, we present radial profiles for the density (top left), enclosed mass (top right), temperature (bottom left), and $\HH$ abundance (bottom right). The spikes in the profiles are an indication of the turbulent morphology at scales $\gtrsim 1$~kpc. As the gas collapses into the DM halo, the gas temperature increases to $\simeq 8000$ K due to shock-heating, slightly below the virial temperature (dashed line). In the central 100 pc of the halo, the collapse is nearly isothermal due to $\HM$ cooling. The evolution is then well-described by the Larson-Penston solution for an isothermal, self-gravitating gas cloud \citep{Larson_69, Penston_69}, and hence the density profile follows the relation $\nh \propto r^{-2}$, as shown by the dashed line in the top left panel. During this period, the $\HH$ abundance, which had increased from $\simeq 10^{-17}$ to $\simeq 10^{-10}$ during the initial collapse, slightly drops to values $\simeq 10^{-11}-10^{-10}$ due to the strong LW background radiation and collisional dissociation.

\begin{figure}
\begin{center}
\includegraphics[scale=.86]{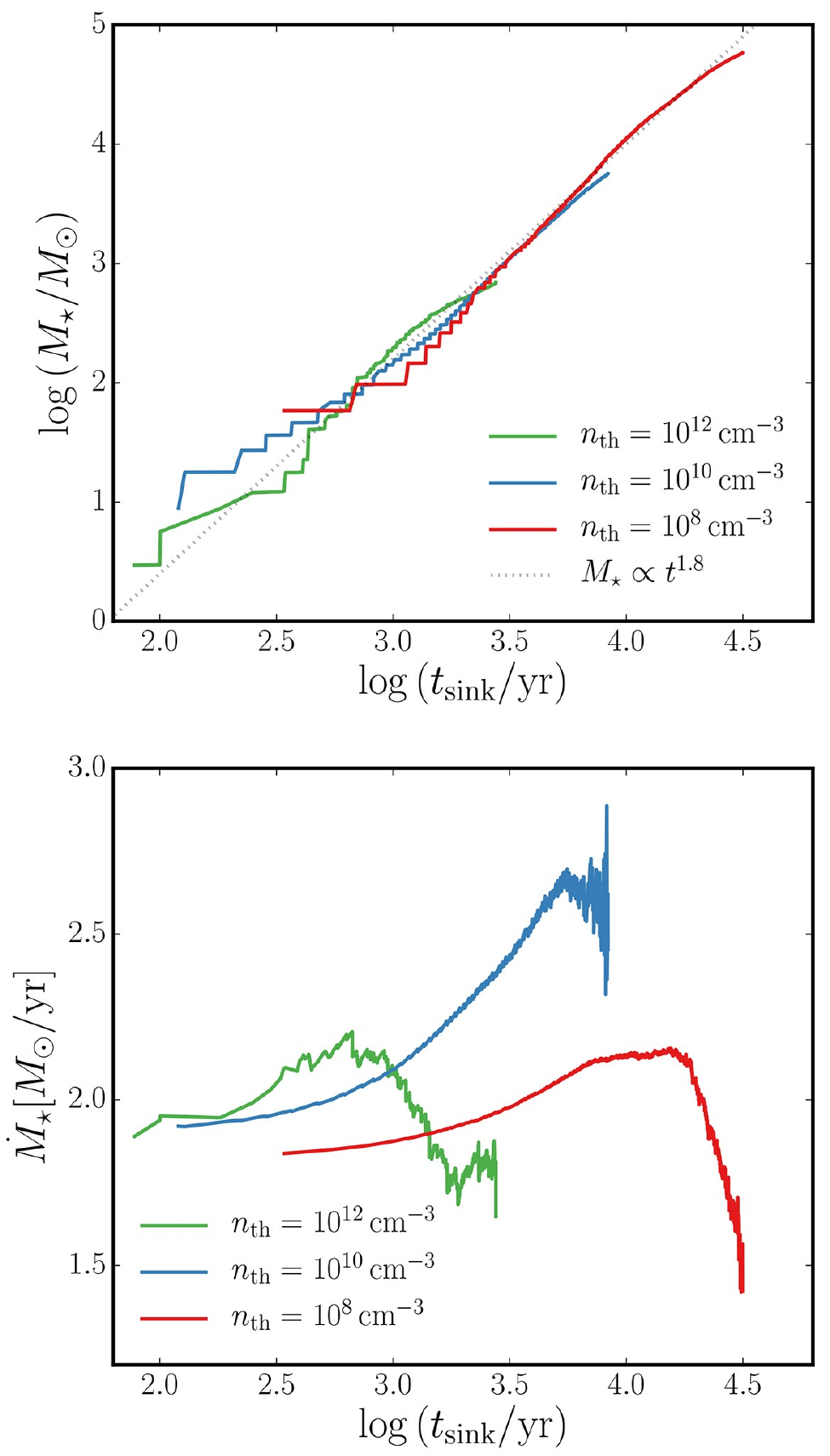}
\caption{Sink mass (top) and accretion rate (bottom) as a function of time since initial sink creation, for threshold densities of $\nth = 10^{8}\,\cmmm$ (red), $\nth = 10^{10}\,\cmmm$ (blue) and $\nth = 10^{12}\,\cmmm$ (green). The sink masses grow from $\mstar \simeq$ 3, 10, and 80 $\msun$ to $\mstar \simeq$ 800, 5000, and 60000 $\msun$, with an average accretion rate $\langle\mstardot\rangle \simeq 2\,\msunyr$ for all threshold densities. This evolution is well-described by the relation $\mstar \propto t_{\rm sink}^{1.8}$ (dotted line), which reflects the initial increase in the accretion rate.}
\label{fig:sinks_evolution}
\end{center}
\end{figure}

\begin{figure}
\begin{center}
\includegraphics[scale=.86]{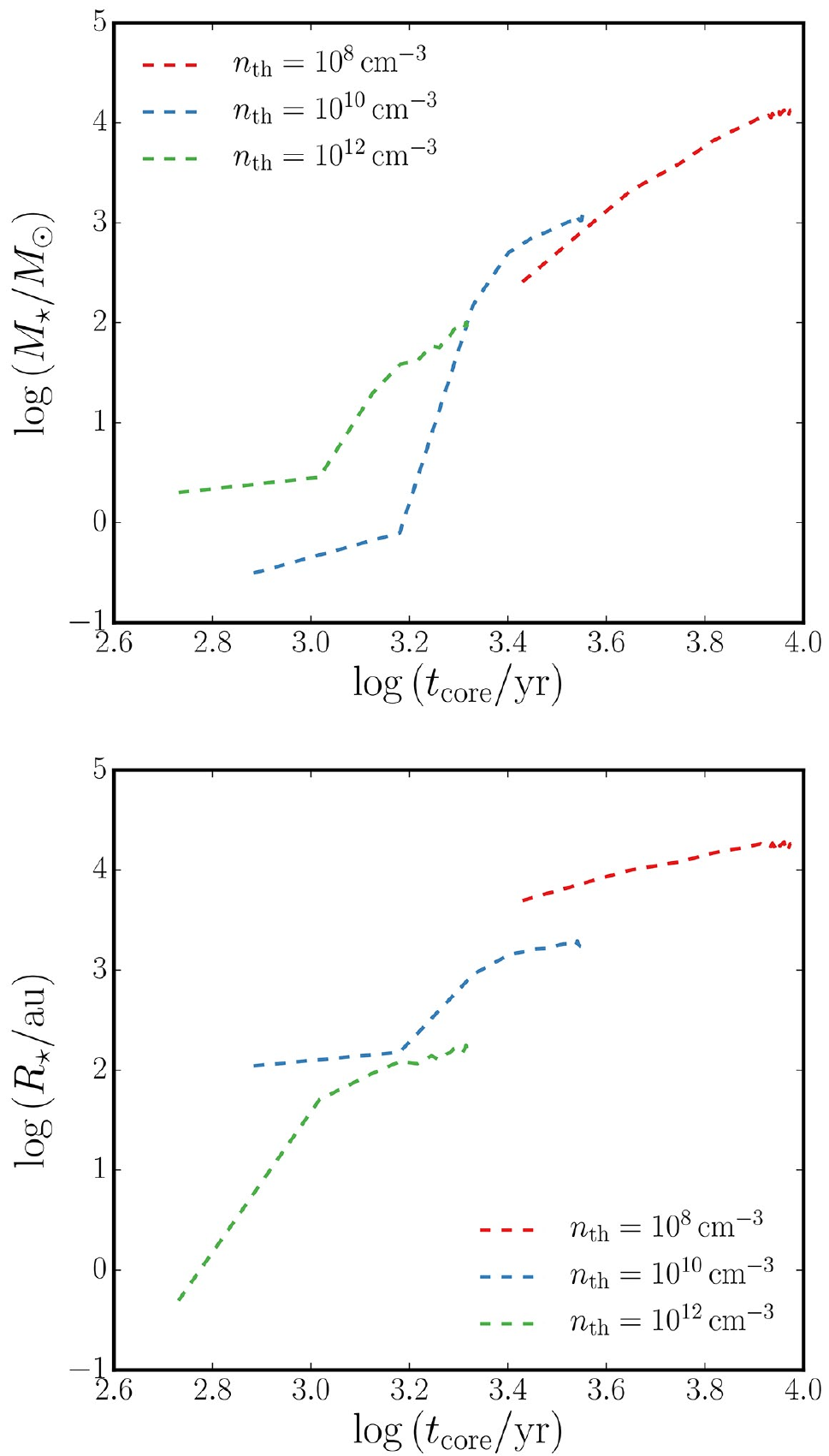}
\caption{Hydrostatic core mass (top) and radius (bottom) as a function of time for threshold densities of $\nth = 10^{8}\,\cmmm$ (red), $\nth = 10^{10}\,\cmmm$ (blue) and $\nth = 10^{12}\,\cmmm$ (green). In the case of the blue and green lines, the mass is characterized by a smooth increase during the early stages, followed by a steeper trend, and a subsequent slower growth rate, while the red line shows a steady increase at all times. The behavior of the $\nth = 10^{8}\,\cmmm$ and $\nth = 10^{10}\,\cmmm$ lines is similar for the evolution of mass and radius, while $\nth = 10^{12}\,\cmmm$ exhibits a steep increase from the beginning, which then changes to a slower pace. At the end of the simulation, the mass and radius of the cores finally reach $\mstar \simeq 10^2$, $10^3$, $10^4\,\msun$ and  $\rstar \simeq 100$, 1,600, 16,500~au, for $\nth = 10^{12}$, $10^{10}$, and $10^{8}\,\cmmm$, respectively.}
\label{fig:cores_evolution}
\end{center}
\end{figure}

\subsection{Central object formation and evolution}
\label{subsec:object_formation}

The central object is formed as soon as the highest density cell surpasses the threshold density $\nth$. In the case of the sink particle simulations, this is defined by the formation and merging of sinks, while in the case of the core simulations the central object is composed by all the gas cells inside the density contour with $\nh = \nth$. Figures \ref{fig:nh_snapshots} and \ref{fig:temp_snapshots} show the density and temperature morphologies, respectively, for sink (top) and core simulations (bottom) at the end of the runs. Columns correspond to threshold densities of $\nth = 10^8$ (left), $10^{10}$ (middle), and $10^{12}\,\cmmm$ (right), and box sizes of 3, 0.3, and 0.03 pc, respectively. It is worth noting that, since we are plotting different times, a direct comparison between the properties of different simulations at those times is not possible. For a convergence study between simulations with different threshold densities and different central object modeling, we refer the reader to Appendix~\ref{app:convergence}.

\begin{figure}
\begin{center}
\includegraphics[scale=.5]{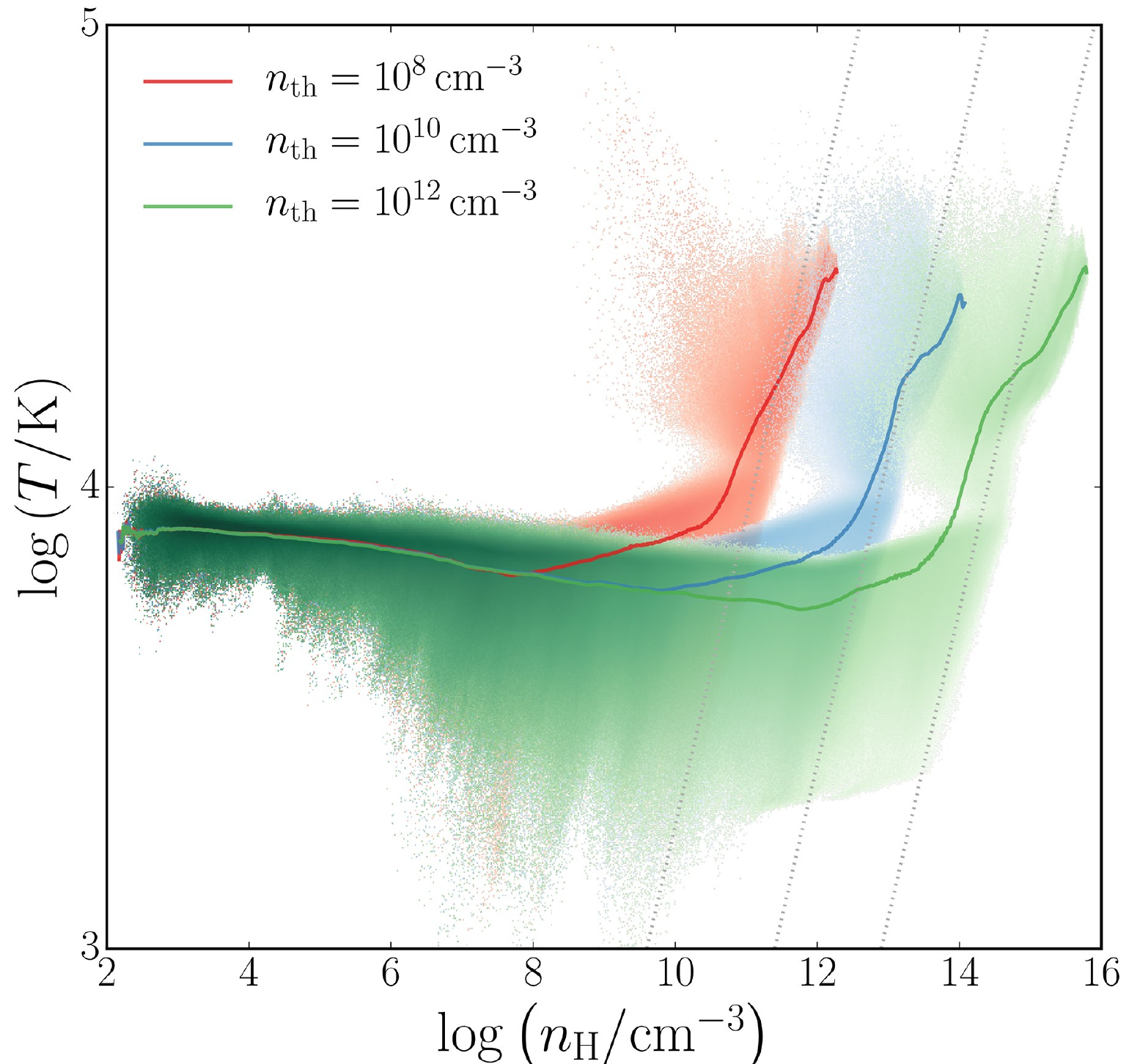}
\caption{Thermodynamics of collapsing gas. Shown is the gas temperature as a function of hydrogen number density for core simulations with threshold densities $\nth = 10^{8}\,\cmmm$ (red), $\nth = 10^{10}\,\cmmm$ (blue) and $\nth = 10^{12}\,\cmmm$ (green). The color-coding indicates mass fractions, from light (lowest) to dark (highest). All simulations show a nearly isothermal evolution until the threshold density is reached. Afterwards, the collapse enters an intermediate stage, where the gas experience a slight increase in temperature over $\simeq$ 2 orders of magnitude in density. Finally, for higher densities, the core follows an adiabatic evolution characterized by the relation $T \propto \nh^{2/3}$ (black dotted lines).}
\label{fig:pspace}
\end{center}
\end{figure}

The gas morphology around the central object varies for different $\nth$. On the left, we can see a disk-like feature of radius $\lesssim 0.5$ pc, while at smaller scales (or, equivalently, higher $\nth$) the gas forms a more elongated structure. In the case of the sink particle simulations, we can discern low-density voids around the sinks, which are caused by gas removal due to sink formation and merging inside the accretion radius. This is not seen in the core simulations, where gas is allowed to keep collapsing, such that some gas cells reach values larger than $\nth$ (yellow pixels in bottom row of Figure \ref{fig:nh_snapshots}). In that case, gas undergoes an adiabatic evolution inside the core and its temperature increases towards the center of the object, visible as a bright dot in the bottom panels of Figure \ref{fig:temp_snapshots}. A similar but more extended increase in temperature is evident in the sink simulations. In contrast to the core cases, this is due to dynamical interaction between sinks that stir and heat up the surrounding gas, reaching values of $T \simeq 10^{4.4}\,{\rm K}$ at scales $\lesssim 0.01\,{\rm pc}$ in the top right panel of Figure \ref{fig:nh_snapshots}. A similar dynamical effect in the presence of sink particles was seen in simulations of Pop~III protostars \citep[][]{Stacy_2010}.

A more quantitative analysis of the evolution of the central object is presented in Figures \ref{fig:sinks_evolution} and \ref{fig:cores_evolution}. In the former, we plot the mass (top) and accretion rates (bottom) of the most massive sink particle as a function of the time after sink creation for $\nth = 10^8$ (red), $10^{10}$ (blue), and $10^{12}\,\cmmm$ (green), while the latter shows the mass (top) and radius (bottom) of the cores (based on the definitions in Section \ref{subsubsec:core}), as a function of core age for the same threshold densities. The initial sink masses are $\simeq$ 3, 10, and 80 $\msun$ from larger to smaller threshold density. This is expected, since these values are determined by the gas mass of the progenitor cells, which are smaller for larger $\nth$, and hence higher resolution. We note that our resolution is not sufficient to capture the formation of protostars at the opacity limit \citep[e.g.][]{Becerra_2017}. However, the long-term evolution, studied here, will not be affected by our idealized treatment of the early growth. At the beginning of the runs, all sink particles have accretion rates $\simeq 2.0\,\msunyr$, which subsequently increase somewhat, until reaching a maximum of $\simeq$ 2.1, 2.6, and 2.2 $\msunyr$ for $\nth = 10^8$, $10^{10}$, and $10^{12}\,\cmmm$, respectively. Eventually, most of the gas around the sinks is accreted and their $\mstardot$ decays to values $\simeq$ 1, 2.4, and 1.5 $\msunyr$. By the end of the runs, their masses have grown to $\simeq$ 800, 5000, and 60,000~$\msun$, with an average accretion rate of $\langle\mstardot\rangle \simeq 2\,\msunyr$. Furthermore, their mass growth is well-described as a function of age by the relation $\mstar \propto t_{\rm sink}^{1.8}$, as indicated by the dotted line in the top panel of Figure \ref{fig:sinks_evolution}. This relation is consistent with the variable accretion rate that we observe in our simulations, which is given by a relation of the kind $\mstardot \propto t_{\rm sink}^{0.8}$. To first order, this increase in the accretion rate as a function of time can be explained considering that it scales with temperature according to $\mstardot \propto T^{3/2}$. As time goes by, the radius of the sink increases, enclosing hotter and hotter gas and resulting in higher values of $\mstardot$. In other words, the Larson-Penston rarefaction wave reaches slightly hotter regions farther out in the envelope towards later times resulting in an increase of the accretion rate. The specific details of how we obtain the 1.8 exponent in that relation are still undetermined and will require a better understanding of the accretion process in future work.

In contrast, the mass of the cores quickly grows during the early stages of their evolution to values $\simeq$ 30, 500, and 2000 $\msun$ for $\nth = 10^{12}$, $10^{10}$, and $10^{8}\,\cmmm$, respectively. From then on, they steadily acquire additional mass until reaching $\mstar \simeq 10^2$, $10^3$, and $10^4\,\msun$ at the end of the runs. A similar trend is observed for the radius of the core, which corresponds to values $\simeq 100$, 1,600, and 16,500~au, once the simulations are terminated. The initial rapid evolution of mass and radius corresponds to a stage right after the highest density cell reaches $\nth$ and before the core enters its adiabatic evolution. This phase occurs between densities $10^8-10^{10}$,  $10^{10}-10^{12}$,  and $10^{12}-10^{13}\,\cmmm$, from smaller to larger threshold density, respectively, and can be distinguished in the phase space diagram of Figure \ref{fig:pspace}. Subsequently, the core enters the adiabatic, slowed contraction phase, characterized by the relation $T \propto \nh^{2/3}$, as shown by the dotted lines. The adiabatic evolution coincides with the slower increase in both mass and radius. On average, the cores have accretion rates around $\simeq 0.8 - 1.4\,\msunyr$, which is smaller than the ones reported for sink particles. This might be due to the definition of the core radius, which is significantly smaller than the accretion radius employed for the sink particles.

\begin{figure}
\begin{center}
\includegraphics[scale=.48]{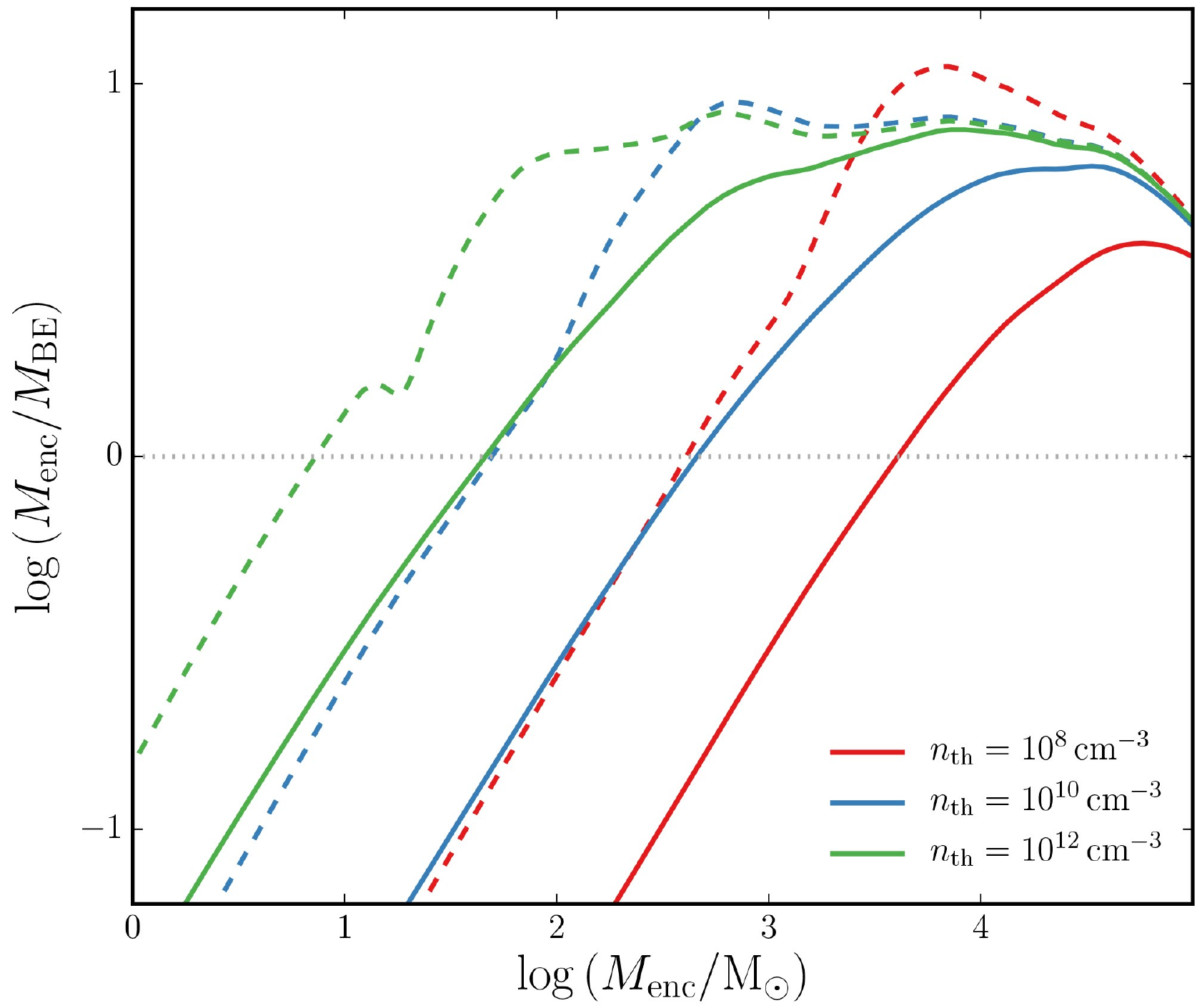}
\caption{Enclosed gas mass over mass-weighted average BE mass as a function of enclosed mass for hydrostatic cores, as soon as they reach the threshold density (solid lines) and at the end of the simulations (dashed lines). Initially, this ratio exceeds unity at $\menc \simeq$ 2500, 250, and 30 $\msun$ for $\nth = 10^{8}\,\cmmm$ (red), $\nth = 10^{10}\,\cmmm$ (blue) and $\nth = 10^{12}\,\cmmm$ (green), which is consistent with the mass of the core before entering the adiabatic phase. Then, as the temperature of the core increases, the point where $\menc/\mbe = 1$ moves to smaller values of the enclosed gas mass until it reaches $\menc \simeq$ 6, 30, 250 $\msun$ at the end of the runs. Some sub-fragmentation is therefore possible towards the later stages of the simulations.}
\label{fig:mbe}
\end{center}
\end{figure}

\subsection{Fragmentation mass scale}
\label{subsec:fragment_mass}

In the case of the core simulations, we can estimate the mass of the collapsed object by calculating the Bonnor-Ebert \citep[BE;][]{Ebert_1955, Bonnor_1956} mass, which is defined as
\begin{equation}
\mbe \simeq 15\,\msun\left( \frac{\nh}{\cmmm} \right)^{-1/2}\left(\frac{T}{\rm K}\right)^{3/2} \mu^{-3/2}\gamma^2,
\label{eq:mbe}
\end{equation}
where $\nh$ is the hydrogen number density, $T$ the temperature, $\mu$ the mean molecular weight, and $\gamma$ the polytropic index. To evaluate this quantity as a function of radius, we calculate the mass-weighted average among cells within a given spherical shell, centered on the highest density cell. In Figure~\ref{fig:mbe}, we show the ratio of enclosed gas mass to BE mass as a function of enclosed gas mass, for two moments: right after the highest density cell has reached the density threshold (solid), and at the end of the runs (dashed), for $\nth = 10^8$ (red), $10^{10}$ (blue), and $10^{12}\,\cmmm$ (green).

After core formation, the enclosed gas mass surpasses the BE value at $\menc \simeq$ 2500, 250, and 30 $\msun$, from lower to higher threshold density, which agrees well with the core mass before entering the adiabatic phase, as illustrated in Figure~\ref{fig:cores_evolution}. As the gas keeps collapsing the temperature increases adiabatically, which is translated into an increase of the BE mass and hence a decrease of the ratio $\menc/\mbe$. As a result, the point at which this ratio exceeds unity is shifted to $\menc \simeq$ 6, 30, 250 $\msun$ at the end of the runs, respectively. This suggests that gas at the center of the halo could fragment into multiple clumps with characteristic masses around those values. Such sub-fragmentation is indeed seen in our simulations, but any fragments are short-lived, and are quickly swept up by frictional forces into the central core \citep[see][]{Hirano_2017}.

\subsection{Mass infall rate}
\label{subsec:mass_infall}

\begin{figure}
\begin{center}
\includegraphics[scale=.5]{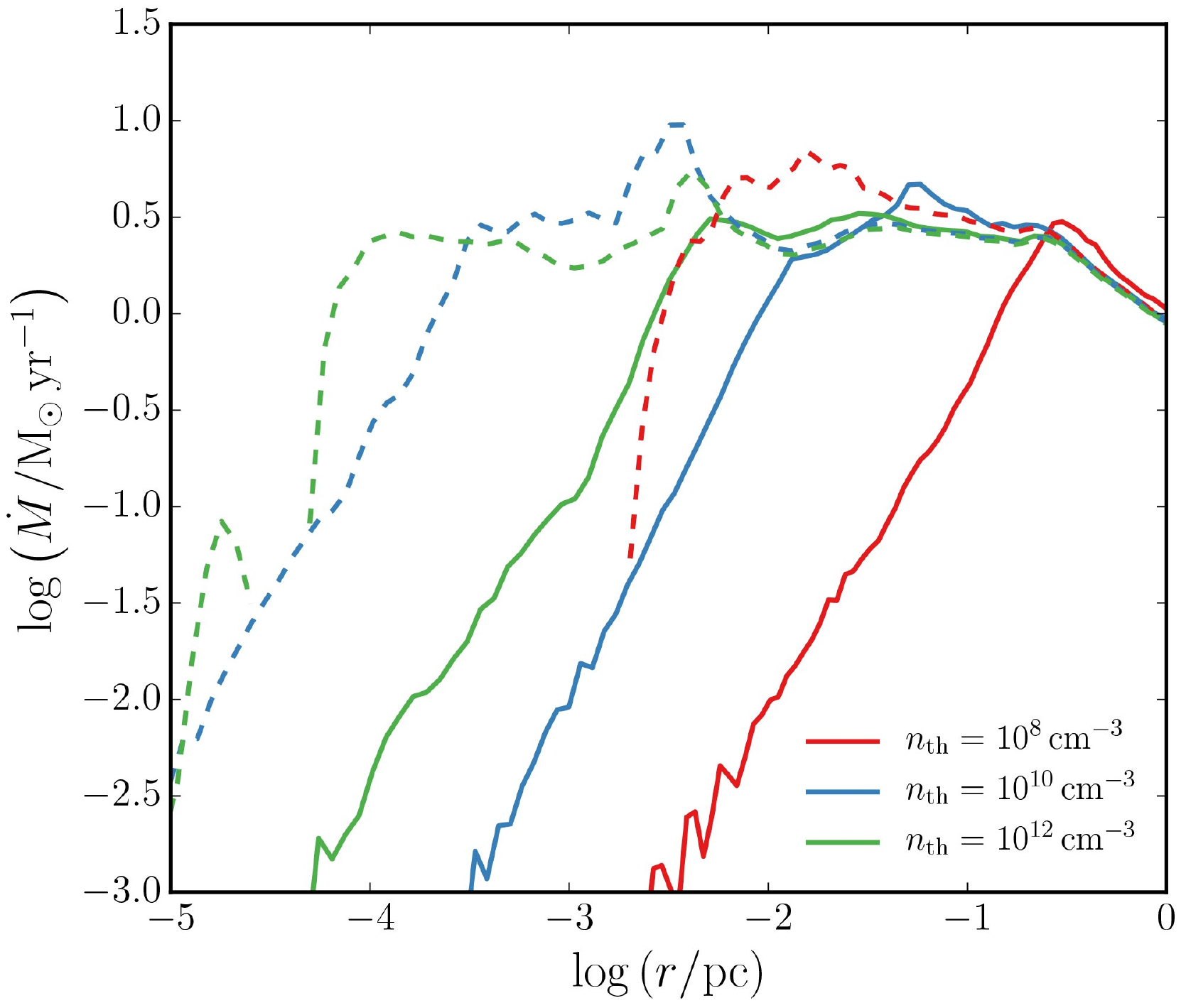}
\caption{Mass infall rate as a function of radius at the end of the runs for sink (solid) and core (dashed) simulations, and threshold densities of $\nth = 10^{8}\,\cmmm$ (red), $\nth = 10^{10}\,\cmmm$ (blue) and $\nth = 10^{12}\,\cmmm$ (green). At large scales, the mass infall rate remains roughly constant at $\mdot_{\rm infall} \simeq 2\,\msunyr$, with a significant drop towards smaller radii to values $\lesssim 10^{-3}\,\msunyr$. In the case of sink simulations, this decrease occurs at larger scales than in the case of core simulations, and it is caused by the removal of gas cells due to accretion onto sinks.}
\label{fig:infall}
\end{center}
\end{figure}

To further study how gas is fed into the central object, we analyze the mass infall rate as a function of radius, which can be calculated as
\begin{equation}
\mdot_{\rm infall}= -4\pi r^2 \rho \vrad,
\end{equation}
where $r$ is the radius, $\rho$ the volume density, and $\vrad$ the radial velocity. In Figure~\ref{fig:infall}, we show the radial profiles for this quantity at the end of the runs, for sink (solid) and core (dashed) simulations and threshold densities of $\nth = 10^8$ (red), $10^{10}$ (blue), and $10^{12}\,\cmmm$ (green).

In the case of sink simulations (solid lines), the mass infall rate reaches a peak of $\simeq 2-3\,\msunyr$ at $r \simeq$ 0.2, 0.02, and 0.003 pc from lower to higher threshold density, while at smaller scales the infall rate drastically falls to values $\lesssim 10^{-3}\,\msunyr$. This decrease is due to the drop in density caused by accretion and merging of sink particles, which constantly removes gas cells inside the accretion radius of the sink particle. In contrast, the core simulations (dashed lines) do not suffer from this limitation and we are able to resolve smaller scales. In such case, the value of the infall rate remains roughly constant around $\mdot_{\rm infall} \simeq 2\,\msunyr$ for all simulations. These values are consistent with the average accretion rates calculated for the sink particles in Section \ref{subsec:object_formation}.

\begin{figure*}
\begin{center}
\includegraphics[scale=1]{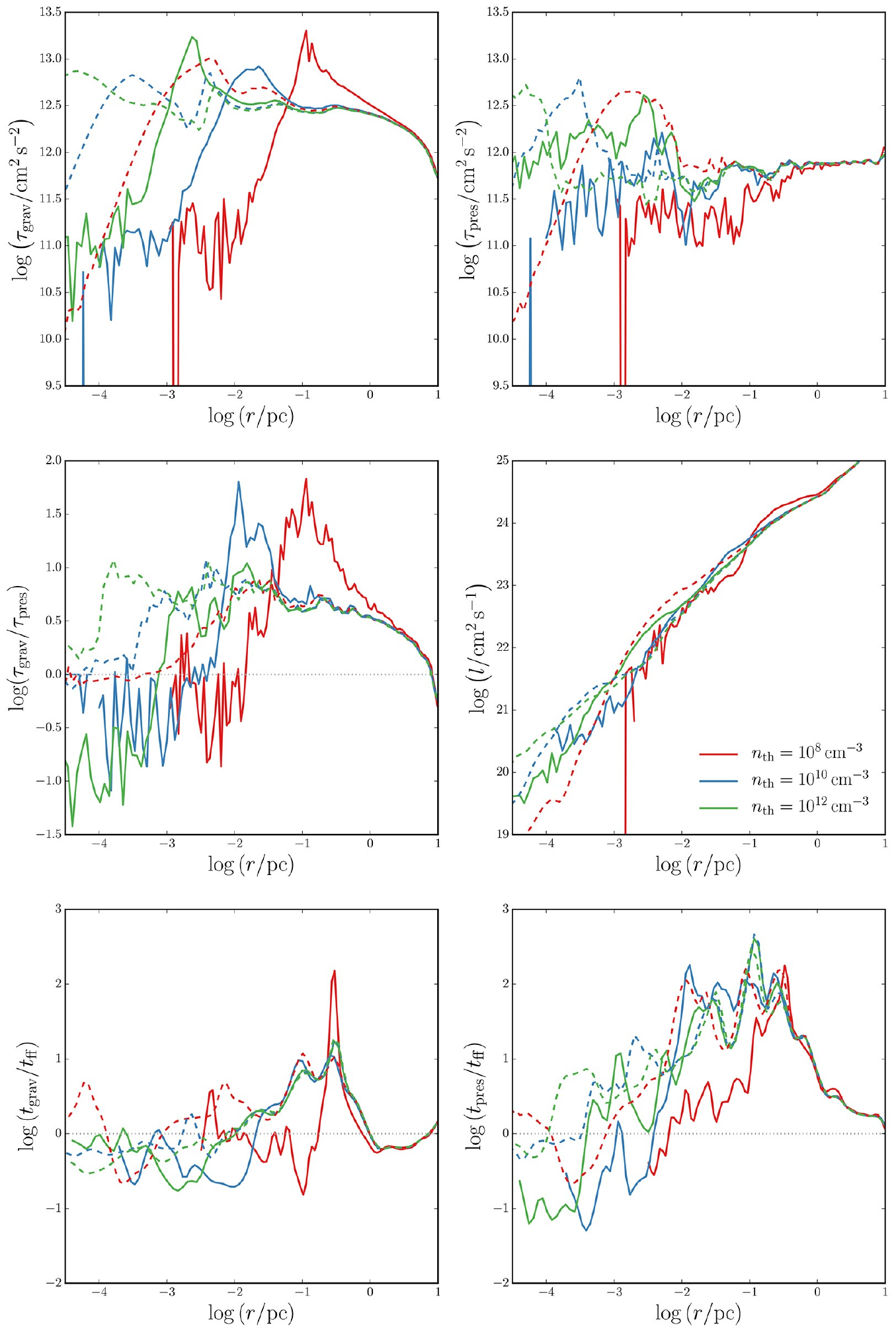}
\caption{Physics of angular momentum transport. Radial profiles of the gravitational torques (top left), pressure gradient torques (top right), their ratio $\taug/\taup$ (middle left), specific angular momentum (middle right), and the timescales for transport driven by $\taug$ (bottom left) and $\taup$ (bottom right). Timescales are normalized to the local free-fall time, $t_{\rm ff}$. The lines represent the values at the end of the runs for sink (solid) and core simulations (dashed) for $\nth = 10^{8}\,\cmmm$ (red), $\nth = 10^{10}\,\cmmm$ (blue) and $\nth = 10^{12}\,\cmmm$ (green). Gravitational torques dominate over pressure gradient torques at almost all scales, except in the inner regions of the sink simulations where gravity is not well-resolved due to accretion of gas cells. These torques are responsible for the removal of angular momentum, which allows the gas to collapse onto the central object.}
\label{fig:torques_sink}
\end{center}
\end{figure*}

\subsection{Angular momentum transport}
\label{subsec:angular_momentum}

One of the key effects that determine the mass that is accreted onto the central protostar is the redistribution of angular momentum in the system. If angular momentum were conserved during the initial collapse, we would expect the gas to form a disk that feeds the object at the center. This disk would be unstable against self-gravity due to the high accretion rates and might fragment into multiple clumps. Angular momentum re-distribution induced by these clumps can drive the evolution of the disk and lead to the formation of a supermassive object \citep{Lodato_2006}. If we consider that gas, on the scale of the host halo, has a similar spin parameter ($\lambda\simeq 0.05$) than DM, we can deduce $\lambda \sim R_{\rm disk}/\rvir$ \citep{Mo_1998}, from which the expected radius for the disk would be $R_{\rm disk} \simeq 50\,{\rm pc}$. Since we do not observe this in our simulations, we can deduce that angular momentum is being removed by gravitational ($\taug$) and pressure ($\taup$) torques on the gas. To quantify the magnitude of the torques, we start by calculating the total baryonic angular momentum at the virial radius, $J_{\rm tot} \simeq 7\times 10^{67}\,{\rm g}\,{\rm cm}^2 {\rm s}^{-1}$, for a virial mass of $\simeq 1.5\times10^{8}\,\msun$. This quantity needs to be removed in the characteristic timescale of collapse, the free-fall time, $t_{\rm ff}\lesssim 10^8$\,yr. Then, we can estimate the torques needed to remove the total angular momentum as $\tau_{\rm tot} \simeq J_{\rm tot}/\tff \simeq 10^{52}\,{\rm g~cm}^2{\rm s}^{-2}$, where ${\boldsymbol \tau}_{\rm tot} = \taug + \taup$. In terms of required torque per mass, this corresponds to values $\gtrsim 10^{12}\,{\rm cm}^2{\rm s}^{-2}$. Let us now assess whether torques of such magnitude are realized in our host haloes.

Previous studies have analyzed how re-distribution of angular momentum affects the evolution of the first stars in the context of minihaloes \citep[e.g.][]{Greif_2012, Hirano_2018}. In order to analyze this for atomic cooling haloes, we follow a similar approach, and examine the torques acting on the gas around the supermassive black hole seed. Specifically, we focus on two qualitatively different kind of torques: gravitational and pressure gradient, which are defined as:
\begin{equation}
\taug = {1 \over \sum_i m_i} \sum_i {\boldsymbol r}_i \times (m_i {\boldsymbol a}_i),
\end{equation}
\begin{equation}
\taup = {1 \over \sum_i m_i} \sum_i {\boldsymbol r}_i \times \left(m_i {\nabla P_i \over \rho_i}\right),
\end{equation}
where $i$ is the index of the cell, ${\boldsymbol r}_i$ its distance to the sink particle or the highest density cell, $m_i$ its mass, ${\boldsymbol a}_i$ its gravitational acceleration, $\rho_i$ its volume density, and $\nabla P_i$ its pressure gradient. Since we are dividing by the factor $\sum_i m_i$, these quantities represent the torques per unit mass.

The top panels of Figure \ref{fig:torques_sink} show the gravitational (left) and pressure gradient torque (right), as a function of radius, at the end of the sink (solid) and core (dashed) simulations, for density thresholds of $\nth = 10^8$ (red), $10^{10}$ (blue), and $10^{12}\,\cmmm$ (green). At large scales, the gravitational torque varies between $10^{12}$ and $10^{13}\,{\rm cm}^2\,{\rm s}^{-2}$ for all simulations. Similar to the behavior of the mass infall rate, torques decline towards smaller scales to values $\lesssim 10^{11.5}\,{\rm cm}^2\,{\rm s}^{-2}$. For the sink simulations, this occurs at larger radii compared to the core case, which is due to the removal of gas cells inside the accretion radius. In contrast, the pressure gradient torques remain roughly constant around $\simeq 10^{12}\,{\rm cm}^2\,{\rm s}^{-2}$ for $r \gtrsim 10^{-2}$\,pc, and experience a small increase to $\simeq 10^{12}-10^{12.5}\,{\rm cm}^2\,{\rm s}^{-2}$ at smaller scales. For ease of comparison, we plot the ratio $\taug/\taup$ as a function of radius (left middle panel of Fig.~\ref{fig:torques_sink}). From there, it is evident that gravitational torques dominate during the collapse of the cloud at both large and small scales. Sink and core simulations differ most strongly at smaller scales, where $\taug$ is not fully resolved because of the constant removal of gas cells due to accretion.

\begin{figure}
\begin{center}
\includegraphics[scale=.62]{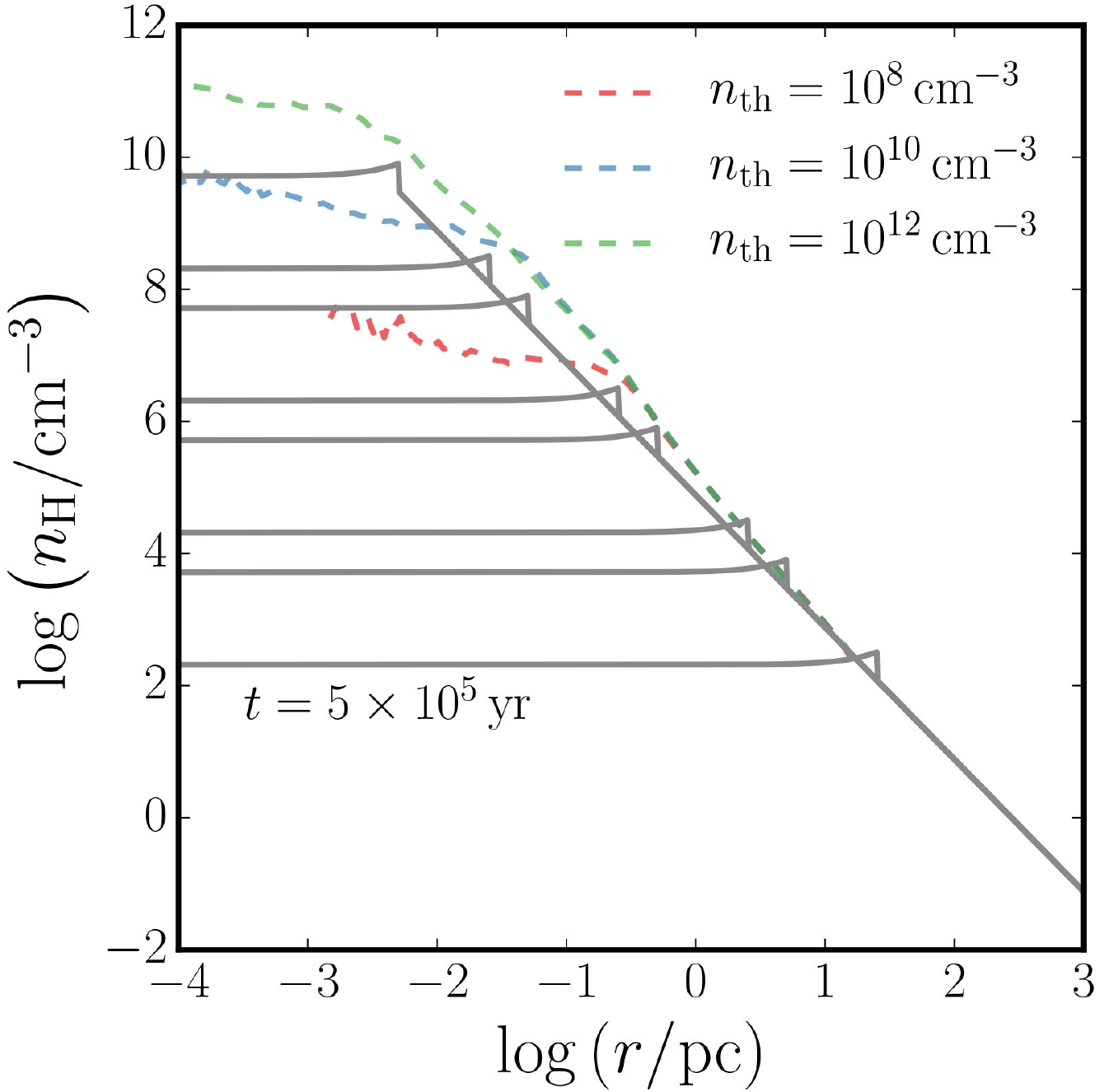}
\caption{Density profile as a function of time for the self-similar solution of a champagne flow \citep{Shu_2002}. From top to bottom, dark grey lines represent $\nh$ at times $t = 10^2$, $5\times10^2$, $10^3$, $5\times10^3$, $10^4$, $5\times10^4$, $10^5$, and $5\times10^5$ yr. The profile is characterized by a nearly flat core that transitions to an isothermal density profile outside the shock radius. The latter is a good approximation for the density profile of our simulations, sufficiently far from a sink (dashed lines).}
\label{fig:density_breakout}
\end{center}
\end{figure}

\begin{figure}
\begin{center}
\includegraphics[scale=.6]{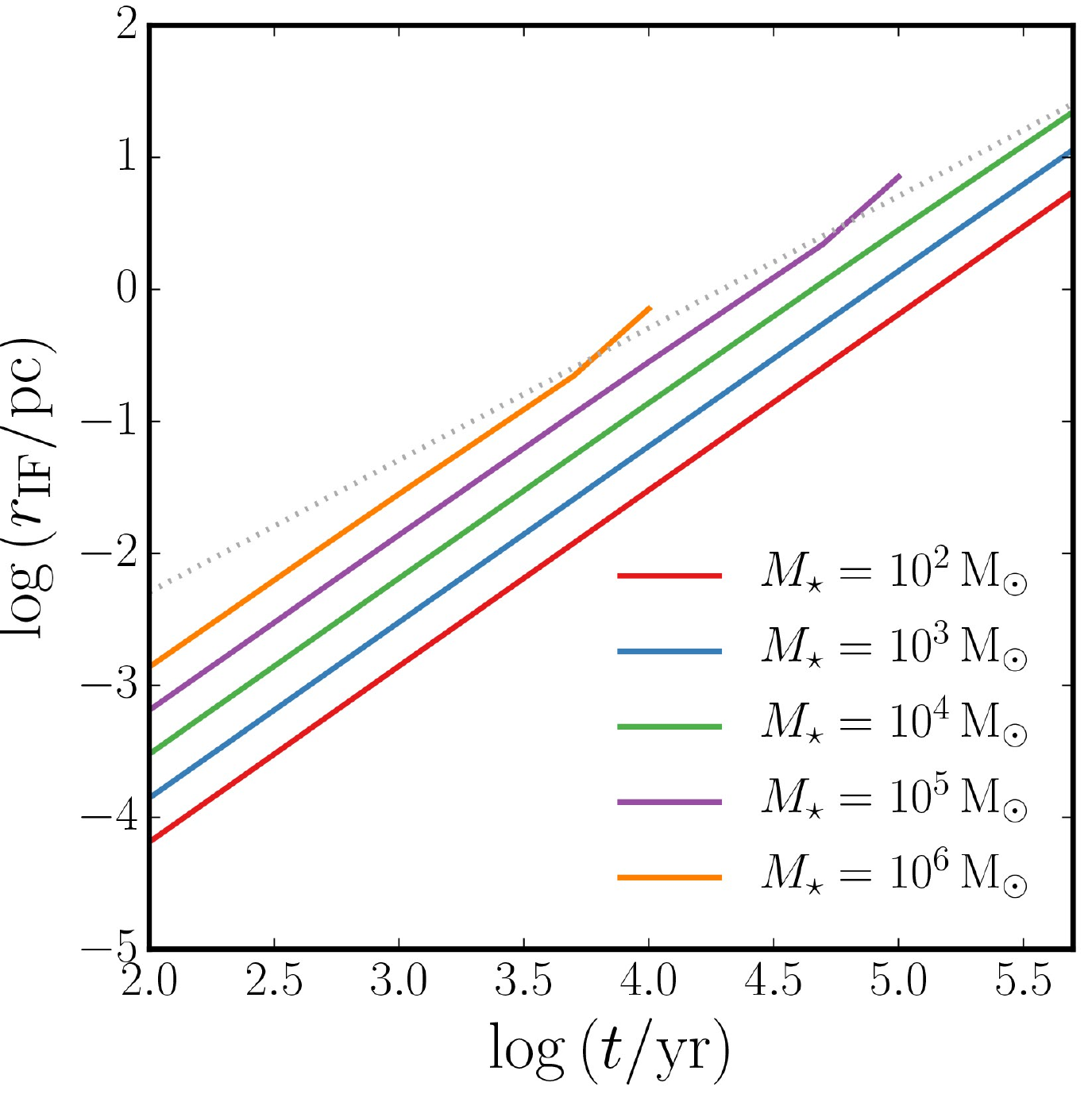}
\caption{Ionization front radius as a function of time for masses $\mstar = 10^2$ (red), $10^3$ (blue), $10^4$ (green), $10^5$ (purple), and $10^6\,\msun$ (orange). $\rIF$ grows from values $\simeq 10^{-3} - 10^{-4}$ pc to $\simeq 3 - 10$ pc after $5\times 10^5$ yr. For masses $\mstar =10^4$ and $10^5\,\msun$, the I-front radius eventually exceeds the shock radius (dotted line), and radiation is able to escape the central core, into the halo and beyond. For smaller masses, radiation break-out has not occurred yet, and radiation remains bottled up.}
\label{fig:rif_vs_time}
\end{center}
\end{figure}

In order to quantify how these torques might affect the redistribution of material around the central object, we analyze their influence on the angular momentum distribution of the collapsing gas. The latter can be calculated as
\begin{equation}
{\boldsymbol l} = {1 \over \sum_i m_i} \sum_i {\boldsymbol r}_i \times (m_i {\boldsymbol v}_i),
\end{equation}
where ${\boldsymbol v}_i$ represents the velocity of the cell with respect to the sink particle or the highest density cell's velocity. We can thus derive the timescales on which each torque operates:
\begin{equation}
\tgrav = {|{\boldsymbol l}|^2 \over {\boldsymbol l} \cdot \taug}, \qquad \tpres = {|{\boldsymbol l}|^2 \over {\boldsymbol l} \cdot \taup}.
\end{equation}
The right middle panel of Figure \ref{fig:torques_sink} shows the specific angular momentum, as a function of the radius, for the same snapshots and simulations indicated above. The steady decrease from $l \simeq 10^{25}\,{\rm cm}^2\,{\rm s}^{-1}$ at $r \simeq 10$ pc  to $l \simeq 10^{20}\,{\rm cm}^2\,{\rm s}^{-1}$ at $r \simeq 10^{-4}$ pc shows that gas is losing its angular momentum as it collapses to the center of the cloud. To explain this, we consider the timescales over which the gravitational (left) and pressure gradient torque (right) act, compared to the free-fall time $\tff = \sqrt{3\pi / 32 G \rho}$ (bottom panels of Fig.~\ref{fig:torques_sink}). In general, $\tgrav \simeq \tff$, except between 0.01 and 1 pc where $\tgrav \simeq 10 \tff$, while in the case of the pressure gradient torques, the ratio $\tpres/\tff$ reaches values $\simeq 100$ in the same region. Gravitational torques, therefore, are more efficient in removing angular momentum, which ultimately impedes the formation of a disk and allows the gas to collapse to the center of the cloud and form a supermassive star.

\subsection{Radiation breakout}
\label{subsec:radiation}

As the central object grows, emission coming from it ionizes the surrounding hydrogen developing an \HII region. The extent of this region is characterized by the location of the ionization front (IF), which at earlier times is bounded to the central star. However, as the central object accretes mass, its emission grows and the \HII region expands from its initial ultra-compact state. A similar early evolution of the developing \HII region has been encountered in radiation-hydrodynamic (RHD) simulations of Pop~III protostar formation \citep[][]{Stacy_2016}. Eventually, the radiation from the fully formed SMBH will break out into the intergalactic medium. Once protostellar radiative feedback becomes important, accretion onto the central object can dramatically change, and the assumptions of our model might not be valid anymore. Here we aim to predict the time of break-out for a supermassive black hole seed by following a similar approach to \citet{Smith_2017}, where the ionization front is modeled as the radius, $\rIF$, within which the recombination rate is in equilibrium with the ionization rate.

\begin{figure*}
\begin{center}
\includegraphics[scale=0.95]{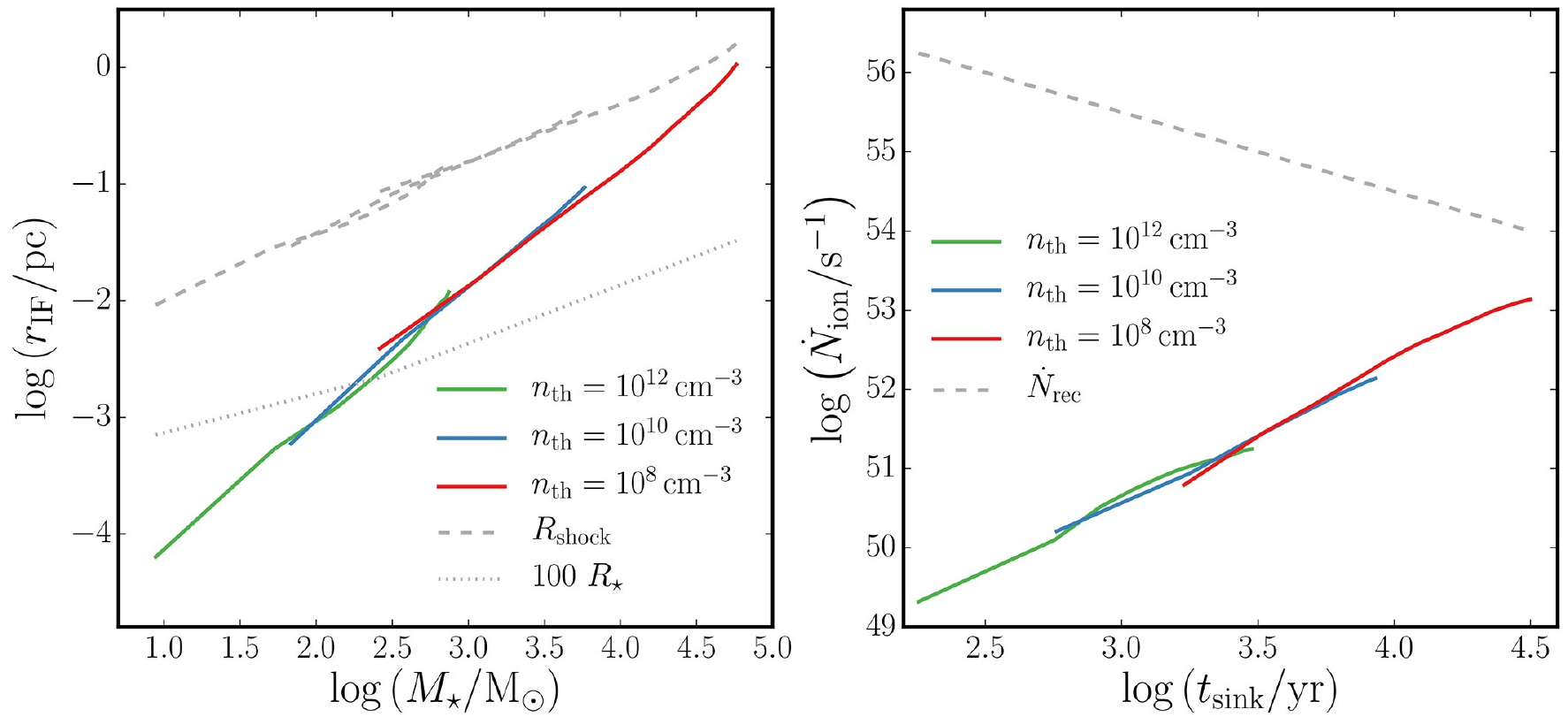}
\caption{Structure of the expanding I-front. Left: Ionization front radius as a function of sink mass based on the data from our simulations with threshold density $\nth = 10^{8}\,\cmmm$ (red), $\nth = 10^{10}\,\cmmm$ (blue) and $\nth = 10^{12}\,\cmmm$ (green). Once the sink mass has reached $\mstar \simeq 5\times10^5\,\msun$, the ionization front radius is of the order of $\simeq 1\,{\rm pc}$ and has not reached the shock radius (dashed lines) yet. Although radiation has not broken out, $\rIF$ has greatly exceeded 100 $\rstar$ (dotted line) and, as a consequence, it has detached from the stellar photosphere. Right: Ionization rates as a function of time after sink creation based on the data from our simulations for the same $\nth$ as in the left-hand panel. After $\simeq 3\times10^4\,{\rm yr}$, $\Nion$ grows to $\simeq 10^{53}\,{\rm s}^{-1}$, which is not sufficient to balance the recombination rate (dashed line). This confirms that radiation has not yet broken out of the host halo by the end of our simulations.}
\label{fig:rif_vs_mass}
\end{center}
\end{figure*}

We begin by writing the recombination rate, as a function of radius $r$, as follows:
\begin{equation}
\dot{N}_{\rm rec} \simeq 4\pi \int_0^{r} \alpha_{\rm B}\nh^2r'^2dr',
\label{eq:recombinations}
\end{equation}
where $\nh$ is the hydrogen number density, $\alpha_{\rm B} = 2.59\times10^{-13}T_4^{-0.7}\,{\rm cm}^3{\rm s}^{-1}$ the effective Case~B recombination coefficient with $T_4 = T/10^4\,{\rm K}$. In principle the density profile can be obtained from the simulations, but it is not modeled accurately inside the sink accretion radius, one problem being that sink particle mass only grows through mergers with other sink particles. Instead, we use the self-similar solution for a champagne flow \citep{Shu_2002}, which assumes a $\rho \propto r^{-2}$ density profile outside a nearly flat inner core. This model has already been used to describe the density evolution in minihaloes \citep[e.g.][]{Alvarez_2006, Wang_2012}, and here we apply the same reasoning to atomic cooling haloes. In such a case, the density profile for the isothermal case with temperature $T_{\rm iso}$ is given by
\begin{equation}
n(r) \simeq 7.7\times10^4 \left(T_{\rm iso}\over 10^4\,{\rm K}\right) \left(r \over 1\,{\rm pc}\right)^{-2}\,\cmmm.
\label{eq:iso_density}
\end{equation}
This profile gives a good description of the evolution of an atomic cooling halo with $T_{\rm iso} \simeq \Tvir \simeq 10^4\,{\rm K}$, as shown in Figure \ref{fig:radial_profiles}.

In contrast, the density profile inside the core is described by
\begin{equation}
{\mh n(r) \over X} = {\alpha(x) \over 4 \pi G t^2},
\label{eq:shu_density}
\end{equation}
where $x=r/\cs t$ corresponds to the similarity variable, $\cs$ to the sound speed of the ionized gas, $t$ to the time after central object formation, $X=0.76$ to the hydrogen mass fraction, and $\alpha(x)$ to a function that characterizes the shape of the density profile in the champagne flow. \citet{Shu_2002} provides a convenient series expansion for this function, using the boundary condition $\alpha = \alpha_0$ at $x=0$:
\begin{equation}
\alpha = \alpha_0 + \frac{\alpha_0}{6} \left( {2\over 3} - \alpha_0 \right) x^2 + \ldots.
\label{eq:alpha}
\end{equation}
The values for $\alpha_0$ can be extrapolated from table~1 in \citet{Shu_2002}, based on the value of $\epsilon \equiv (c_{\rm s, i}/\cs)^2$, where $c_{\rm s, i}$ and $\cs$ are the initial and ionized isothermal sound speeds. Here we have used $T_{\rm i} \simeq 10^4\,{\rm K}$ and $T = 3\times10^4\,{\rm K}$  for the initial and \HII region temperatures, respectively. The transition between both profiles occurs at the shock radius, $r_{\rm sh} = v_{\rm s} t$, where the shock velocity $v_{\rm s} = x_{\rm s} \cs$ can be calculated with the value for $x_{\rm s}$ extracted from the same table as above.

Figure \ref{fig:density_breakout} shows the density profile for the case of an atomic cooling halo in gray lines at times $t = 10^2$, $5\times10^2$, $10^3$, $5\times10^3$, $10^4$, $5\times10^4$, $10^5$, and $5\times10^5$\,yr, from top to bottom. For reference, we have included the density profiles from the last snapshot of each sink simulation (dashed lines) for the different threshold densities $\nth = 10^{8}\,\cmmm$ (red), $10^{10}\,\cmmm$ (blue), and $10^{12}\,\cmmm$ (green). Our model predicts that in response to the photo-heating, the density inside the ionized region continuously decreases as time goes by, reaching values as low as $\simeq 10^2\,\cmmm$ at $t = 5\times10^5\,{\rm yr}$. In contrast, the simulations, not taking into account any RHD effects, do not exhibit such a roughly constant-density inner core. Instead, the density keeps increasing near the center, reaching values a few orders of magnitude larger than our analytical model.

Next, we need to account for the ionizing radiation originating from the central object, evaluating the ionizing photon rate according to
\begin{align}
\Nion &= \frac{\pi L_{\rm Edd}}{\sigma_{\rm SB}T_{\rm eff}^4} \int_{\nu_{\rm min}}^\infty \frac{B_\nu}{h\nu}d\nu\\
&\simeq 2.37 \times 10^{54} {\rm s}^{-1} \left( \mstar \over 10^6\,\msun \right)\mbox{\ ,}
\label{eq:ionizations}
\end{align}
where $\sigma_{\rm SB}$ is the Stefan-Boltzmann constant, $B_\nu$ the Planck function, and $h\nu_{\rm min}$ = 13.6 eV. For simplicity, we here assume a blackbody source with an effective temperature of $T_{\rm eff} = 10^5\,{\rm K}$ emitting at the Eddington luminosity, $L_{\rm Edd} = 4\pi G\mstar\mh c/\sigma_T$, where $\mstar$ corresponds to the mass of the sink particle. Such properties are characteristic for very massive stars \citep[e.g.][]{Bromm_2001}.
Inside the ionization front, the recombination (Equ.~\ref{eq:recombinations}) and ionization (Equ.~\ref{eq:ionizations}) rates are in equilibrium, hence
\begin{equation}
4\pi \int_0^{\rIF} \alpha_{\rm B}\nh^2r^2dr \simeq 2.37 \times 10^{54} {\rm s}^{-1} \left( \mstar \over 10^6\,\msun \right).
\label{eq:equilibrium}
\end{equation}
From this, and using Equations \ref{eq:iso_density} and \ref{eq:shu_density} to describe the density profile, we can estimate the radius of the ionization front, $\rIF$, as a function of time and mass of the star. These results are shown in Figure \ref{fig:rif_vs_time} for masses of $\mstar = 10^2$ (red), $10^3$ (blue), $10^4$ (green), $10^5$ (purple), and $10^6\,\msun$ (orange). The ionization front increases from values $\simeq 10^{-4}-10^{-3}\,{\rm pc}$ at $t \simeq 100\,{\rm yr}$, to $\simeq 1 - 10\,{\rm pc}$ after $5 \times 10^5\,{\rm yr}$. As the value of $\rIF$ grows, it might be able to reach the shock radius, defined in the Shu solution (gray dotted lines). At that point, radiation would be able to escape from the central core, overrunning the halo and reaching the IGM. This occurs at the breakout time $t_{\rm B} \simeq 5\times10^4\,{\rm yr}$ and $5\times10^3\,{\rm yr}$ for masses $\mstar = 10^5\,\msun$ and $10^6\,\msun$, respectively. For lower values of $\mstar$, the ionization front has not reached the shock radius after $5 \times 10^5\,{\rm yr}$, and radiation is still bottled up, but just barely so.

Now, we apply this model to our simulations. In that regard, we extract the sink masses and the times after sink creation from the outputs and insert them into Equation \ref{eq:equilibrium} to estimate $\rIF$. The results are plotted in the left panel of Figure \ref{fig:rif_vs_mass} for threshold densities of $\nh = 10^8$ (red), $10^{10}$ (blue), and $10^{12}\,\cmmm$ (green). The initial I-front is located at a radius $\rIF\simeq 7 \times 10^{-5}\,{\rm pc}$, once the sink is created with $\mstar \simeq 10\,\msun$ in the $\nh = 10^{12}\,\cmmm$ case. As the sink mass grows to $\mstar \simeq 5\times10^5\,\msun$, $\rIF$ increases to $\simeq 1\,{\rm pc}$ but has not yet reached the shock radius (gray dashed lines), which implies that radiation break-out has not happened at the end of our simulations. For reference, we have included $100\rstar$ as a gray dotted line, where $\rstar$ is the radius of the growing protostar, as calculated in \citet{Becerra_2017}. We use this radius as an indication of the point at which radiation detaches from the stellar photosphere, which occurs at $\mstar \simeq 200\,\msun$ for our models. From that moment radiation might influence the gas surrounding the central object and affect its growth rate. In the right-hand panel of the same figure, we use our simulation data to calculate the ionization rates as a function of time, with threshold density as a parameter. Additionally, we plot the recombination rates calculated from Equation~\ref{eq:recombinations}, integrated out to $r = \rvir$ (gray dashed lines). $\Nion$ grows from $\simeq 3\times10^{51}\,{\rm s}^{-1}$ at $t \simeq 200\,{\rm yr}$ to $\simeq 10^{53}\,{\rm s}^{-1}$ after $\simeq 3\times10^4\,{\rm yr}$. Nevertheless, this is not sufficient to balance the recombination rate, $\dot{N}_{\rm rec} \simeq 10^{54}\,{\rm s}^{-1}$. Hence, radiation has not yet broken out from the host halo.

Recently, \citet{Chon_2018} and \citet{Luo_2018} have performed radiation hydrodynamical (RHD) simulations of the collapse of an atomic cooling halo and have followed the evolution of the ionized region around the central protostar. In the appendix of \citet{Luo_2018}, they analyze the size of the expanding \HII region, which starts at $\simeq 4 \times 10^{-4}\,{\rm pc}$ when $t \simeq 10\,{\rm yr}$ and increases to $\simeq 2 \times 10^{-3}\,{\rm pc}$ around $t \simeq 270\,{\rm yr}$. On the other hand, \citet{Chon_2018} reported that once the protostar reaches $\mstar \simeq 6000\,\msun$, the \HII region has a size of $\simeq 400\,{\rm au}$, which suggests that the ionized gas at that point is gravitationally bound and will remain in that state until the mass of the central star exceeds $\simeq 10^5\,\msun$. Around 300 yr after the formation of the protostar, our numerical estimations are consistent with those of \citet{Luo_2018}, while when the mass of the central object reaches $\simeq 6000\,\msun$, our model suggests values about an order of magnitude larger than those quoted by \citet{Chon_2018}. Furthermore, this also implies that within our prescription the initially ultra-compact \HII region has detached from the protostar, while \citet{Chon_2018} suggest that it is still gravitationally bound. On the other hand, \citet{Regan_2018b} estimate the extent of the ionized region to be around $\simeq 3000 \,{\rm au}$, based on  simulations where they include radiative feedback from the first supermassive stars in atomic cooling haloes, consistent with our results. We plan to run fully coupled RHD cosmological simulations in the future to gain a clearer picture of the crucial role of radiation in the evolution of the central protostar.

\section{Caveats}
\label{sec:caveats}

In this work, we have set out to describe the full picture of how a massive protostar of $\simeq 10\,\msun$ becomes a supermassive black hole seed of $\simeq 10^5\,\msun$. In contrast to our previous work in \citet{Becerra_2015}, we have added an improved treatment for $\HM$ continuum cooling up to densities of $\nh \simeq 10^{16}\,\cmmm$, as detailed in \citet{Becerra_2017}. To be able to follow the evolution of the central object for a longer time, we have implemented sink particles. A drawback of this technique is that the density structure inside the sink particle accretion radius is not accurately captured. For the same reason, quantities that depend on the density, such as the mass infall rate and gravitational torques, are not truly resolved within that distance. To account for that and gain a more complete picture of the processes at small scales, we have run simulations with the same threshold densities, but employing a stiff equation of state to model the central object. Both implementations are complementary, providing us with a broader perspective on the evolution of the protostar.

Furthermore, the sink particles used in our study replace gas cells above a certain threshold density and hence do not directly represent stars. Adding more requirements to form a sink, such as converging, Jeans-unstable, and gravitationally bound gas, might affect the evolution of these particles. In such case, to compare the sinks and the artificially-stiffened equation of state approaches, we might need a more precise definition of a core in the latter case. At the same time, a more restrictive recipe would also imply that less gas cells are converted to sink particles, potentially resulting in a lower accretion rate. Regardless of this, both approaches show similar values for the mass and accretion rates of the central object, which are also consistent with results from previous works \citep[e.g.][]{Regan_2009, Latif_2013a, Latif_2013d, Shlosman_2016, Regan_2018}. Therefore, we do not expect this to change our conclusions significantly.

In addition, we have neglected a number of physical processes that might influence the growth of such an object at some stage during its evolution. In particular, we have not modeled the radiation emerging from the central source. For example, previous studies have discussed that Ly$\alpha$ radiation might be trapped on parsec-scales around the BH seed, which would allow the temperature, and hence the Jeans mass, to increase by a factor of a few. This would result in a change of the adiabatic index to $\gamma = 4/3$, which might delay the initial collapse of the halo \citep{Ge_2017}. Eventually, Ly$\alpha$ radiation might also affect the long-term evolution of the central object by shaping its surroundings \citep{Smith_2017}. Both of these studies used post-processing routines in their analysis. In order to understand the full picture of the formation of SMBH seeds, we will need a fully coupled simulation, where Ly$\alpha$ radiation transport is calculated on-the-fly for self-consistent and higher order description of the processes involved. In this context, \citet{Luo_2018, Ardaneh_2018, Chon_2018} have recently presented pioneering RHD simulations of atomic cooling haloes, while \citet{Regan_2018b} have performed simulations including the effects of radiative feedback from the first supermassive stars. We plan to address the impact of radiation on the surrounding gas in a follow-up study, where we will perform fully-coupled cosmological RHD simulations.

Lastly, we have not included the effects of magnetic fields, which might become important in the direct collapse scenario. For example, magnetic pressure might provide additional support against gravity, which delays the formation of a supermassive star \citep{Latif_2014b}. Future work should include a detailed treatment of magneto-hydrodynamic effects in the context of atomic cooling haloes.

\section{Summary and conclusions}
\label{sec:summary}

We have performed a suite of six simulations to study the buildup of SMBH seeds at high redshifts. We follow the formation and evolution of a supermassive protostar from $\mstar \simeq 10\,\msun$ to $\mstar \simeq 10^5\,\msun$. For this purpose, we model the central object using two methods: sink particles and artificially-stiffened hydrostatic cores, considering three different threshold densities, $\nth = 10^8$, $10^{10}$, and $10^{12}\,\cmmm$, to address the late, intermediate, and early stages of the BH seed. The simulations employ a primordial chemistry network that evolves five species (H, $\HH$, $\HM$, $\HP$, and $\e$), and includes $\HH$ line emission, $\HH$ collision-induces emission, Ly$\alpha$ cooling, $\HM$ continuum cooling, as well as inverse Compton cooling. In addition, we have included a uniform LW background radiation field of strength $J_{\rm 21} = 10^5$ to prevent the formation of stars in progenitor minihaloes.

During the initial collapse the gas is shock-heated to a sub-virial temperature $T \simeq 8000$ K. The molecular hydrogen abundance remains very small throughout the collapse, due to an external LW background radiation that photo-dissociates $\HH$ within the halo, and the collisional dissociation at higher density. As a result, atomic hydrogen cooling dominates in the form of Ly$\alpha$ emission up to densities $\nh \simeq 10^6\,\cmmm$, where the gas becomes optically thick to that radiation, and $\HM$ bound-free and free-free emission up to $\nh \simeq 10^{17}\,\cmmm$. Because of this, the gas evolves nearly isothermally over many orders of magnitude in density, which is reflected in a profile of the form $\nh \propto r^{-2}$. Once the highest-density cell reaches a pre-set threshold density, $\nth$, we create the central object, which is represented by a sink particle or by a hydrostatic core where cooling has been artificially suppressed.

By analyzing simulations with three threshold densities $\nth = 10^8$, $10^{10}$, and $10^{12}\,\cmmm$, we follow the evolution of the central object from $\mstar \simeq 10\,\msun$ to $\mstar \simeq 6\times 10^4\,\msun$. Its growth is characterized by a relation of the form $\mstar \propto t_{\rm sink}^{1.8}$, where $t_{\rm sink}$ is the age of the sink particle. Accretion rates are in the narrow range $1- 2.5$\,$\msunyr$, with an average value of $\langle\mstardot\rangle \simeq 2\,\msunyr$, consistent with the mass infall rates for the collapsing gas. This process is mainly driven by strong gravitational torques, $\taug \simeq 10^{12.5}\,{\rm cm}^2\,{\rm s}^{-2}$, acting on timescales comparable to the free-fall time. The initial angular momentum is thus efficiently removed, allowing the gas to feed the central object. As the protostar grows, radiation originating from it ionizes the surrounding material, thus creating an initially ultra-compact, but subsequently expanding \HII region. Towards the end of the lowest-resolution simulation, the rate of ionizing photons, as estimated with our idealized protostellar evolution sub-grid model, is still below the recombination rate. This would imply that the ionizing radiation is still bottled up, and has not yet escaped from the host halo. On the other hand, this emission detaches from the photosphere of the central object when its mass is $\simeq 200\,\msun$. From then on, radiative feedback can influence accretion of gas onto the central object, thus ultimately affecting its final mass. For realistic modeling, the central density structure has to be known. With sinks, however, this structure is not properly resolved at small scales. To approximately assess the situation, we have here used the self-similar solution for a champagne flow to deduce the location of the I-front, and estimate the breakout time. A more accurate description of the radiative feedback from the central source and its effect on the gas requires a fully coupled, on-the-fly implementation of RHD in a cosmological context.

The understanding of how supermassive black holes formed so early in cosmic history is making remarkable progress, based on advances in numerical technology \citep[e.g.][]{Johnson_2016, Latif_2016b, Smith_2017b, Valiante_2017}. Tantalizingly, the early Universe may have provided unique conditions for the accelerated emergence of massive BHs, and it is becoming evident that they have played an important part in shaping early cosmic history. The first SMBHs also provide luminous beacons of the high-redshift Universe, to be probed with the next generation of observational facilities, such as the {\it JWST} and the extremely large telescopes currently being built on the ground. To fully harness the promise of these facilities, simulations are vital in elucidating the key observational signatures.

\section*{Acknowledgements}

We would like to thank the anonymous referee for the constructive comments that helped to improve our paper. VB was supported by NSF grant AST-1413501. The simulations were carried out at the {\em Odyssey} supercomputer in the Faculty of Arts and Sciences at Harvard University.



\bibliographystyle{mnras}


\appendix

\section{Convergence}
\label{app:convergence}

Following our discussion in Section \ref{subsec:object_formation}, we here present evidence that our simulations are converged for different central object models and different threshold densities. For that purpose, we consider number density and temperature projections for sink and stiff equation of state (EOS) simulations, and compare the resulting morphologies at similar times in their evolution. Specifically, in Appendix \ref{subapp:sink_sink} we present a comparison between sink simulations with different $\nth$, while in Appendix \ref{subapp:sink_core} we examine different central object modeling using the same $\nth$.

\begin{figure*}
\begin{center}
\includegraphics[scale=0.95]{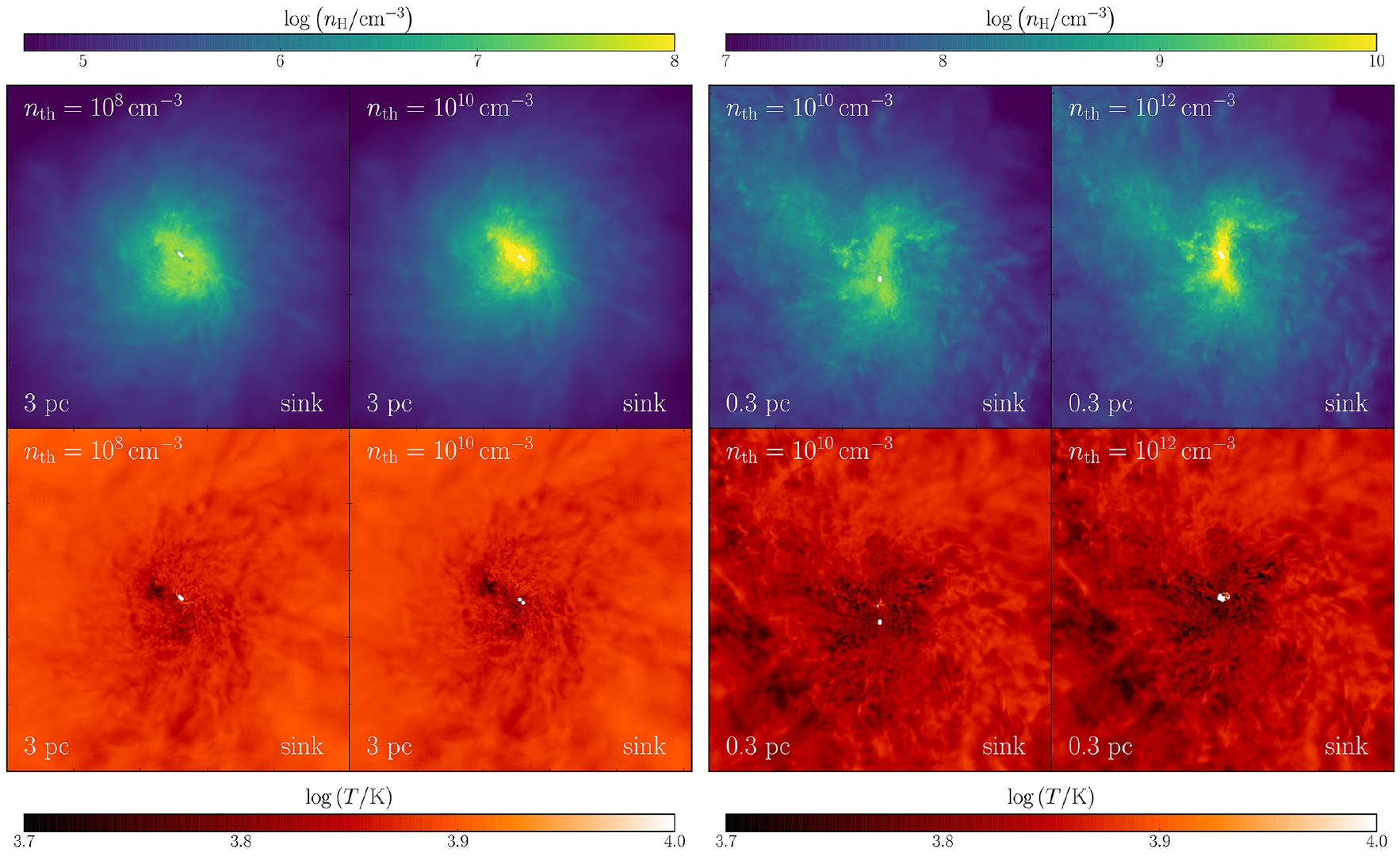}
\caption{Number density (top) and temperature (bottom) projections for sink particle simulations with different threshold densities. Left panel shows a comparison between $\nth = 10^8\,\cmmm$ (left column) and $\nth = 10^{10}\,\cmmm$ (right column) for a box size of 3 pc, while the right panel displays snapshots for $\nth = 10^{10}\,\cmmm$ (left column) and $\nth = 10^{12}\,\cmmm$ (right column) in a 0.3 pc projection. Both density and temperature fields show similar distribution and values, independent of the threshold density used, demonstrating that these simulations are reasonably converged.}
\label{fig:convergence_sinks}
\end{center}
\end{figure*}

\subsection{Sink particles simulations}
\label{subapp:sink_sink}

Figure \ref{fig:convergence_sinks} shows number density (top) and temperature (bottom) projection for sink particle simulations using different threshold densities. The left panel compares a box size of 3 pc for $\nth = 10^8\,\cmmm$ (left column) and $\nth = 10^{10}\,\cmmm$ (right column), while the right panel shows a comparison between $\nth = 10^{10}\,\cmmm$ (left column) and $\nth = 10^{12}\,\cmmm$ (right column) in a 0.3 pc projection. Additionally, we set the upper limit of the color bar to the lowest value of $\nth$ in each panel in order to distinguish all the gas particles that are above the lowest $\nth$ as bright yellow dots in the density projection. In this way, we effectively use bright yellow dots to represent the gas particles in the highest $\nth$ simulation that have been replaced by sink particles and accreted by the central object in the lowest $\nth$ simulation.

In general, we can see that the gas has a similar morphology in both panels, showing a disk-like structure in the left one, while the right one reveals a bar-like pattern at smaller scales. Similarly, the temperature distribution encloses the same range and distribution of values for both comparisons. Nevertheless, from the images we can see that the location of the central sink particle might change slightly when using different threshold densities. This, however, is small compared to the accretion radius and hence should not influence its evolution and growth.

\begin{figure*}
\begin{center}
\includegraphics[scale=0.7]{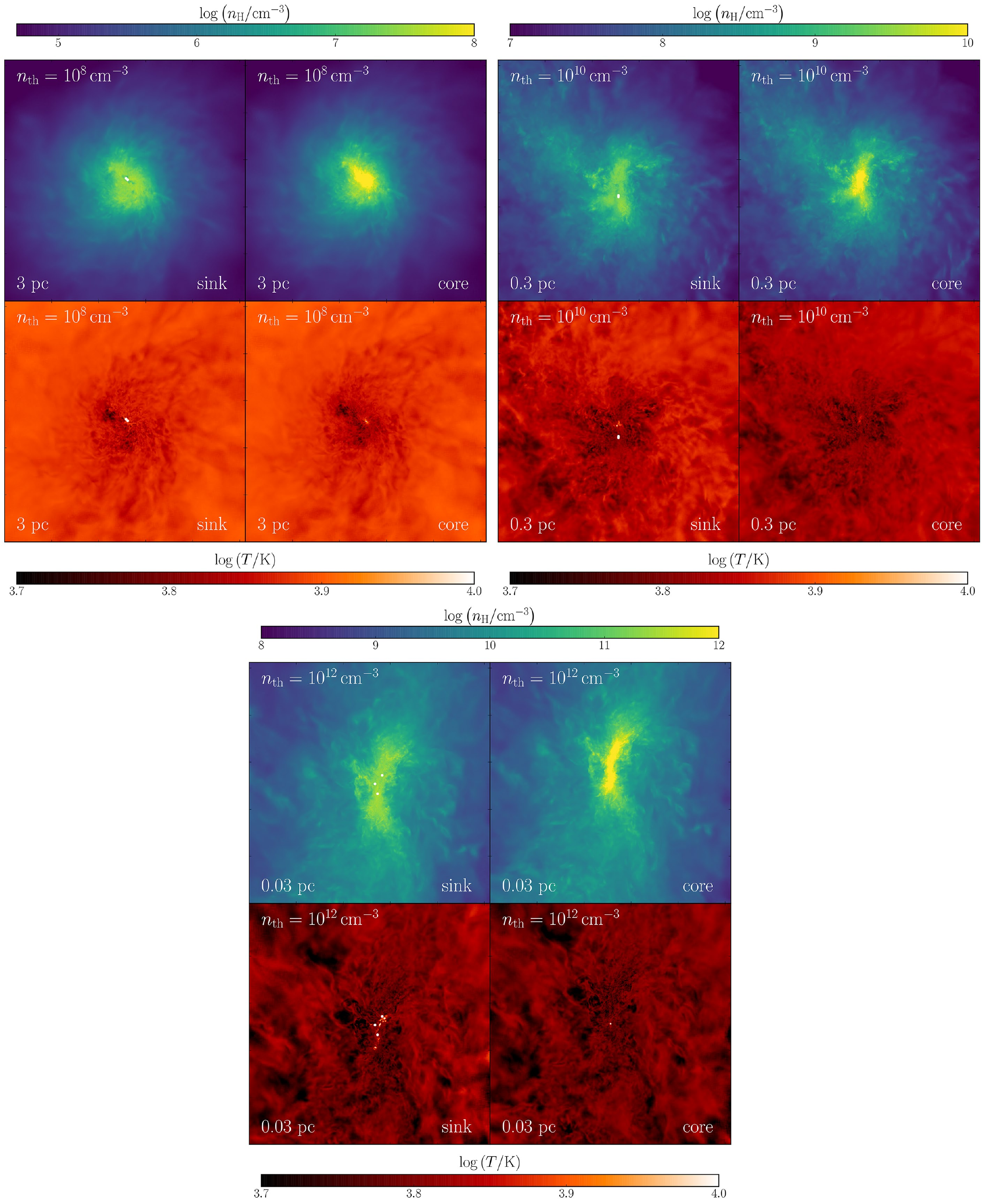}
\caption{Number density (top) and temperature (bottom) projections of sink particles and stiff EOS simulations with the same threshold density. We display snapshots using $\nth = 10^8\,\cmmm$ in a 3 pc box in the top left panel, $\nth = 10^{10}\,\cmmm$ in a 0.3 pc box in the top right panel, and $\nth = 10^{12}\,\cmmm$ in a 0.03 pc in the bottom panel. The simulations show a similar distribution in both density and temperature, independent of the central object modeling approach used. Although we can appreciate small differences in temperature for $\nth = 10^{12}\,\cmmm$ due to strong gravitational interactions between sink particles, this does not impact the evolution of the central object.}
\label{fig:convergence_cores}
\end{center}
\end{figure*}

\subsection{Sink particles and stiff EOS simulations}
\label{subapp:sink_core}

Figure \ref{fig:convergence_cores} presents a similar comparison between number density (top) and temperature (bottom) projections for different central object modeling and using the same threshold density. We show a box size of 3 pc and $\nth = 10^8\,\cmmm$, 0.3 pc and $\nth = 10^{10}\,\cmmm$, and 0.03 pc and $\nth = 10^{12}\,\cmmm$ in the top left, top right, and bottom panels, respectively. For each panel we display the sink particle method on the left column and the stiff EOS approach on the right one. In addition, we set the upper limit of the color bar to the value of $\nth$, hence representing the gas particles above that threshold as bright yellow dots. These gas cells have become sink particles and been accreted by the central object in the left columns, while in the right one they have become part of the hydrostatic core as defined in Section \ref{subsubsec:core}.

Analogous to the previous figure, it is evident that the simulations using the same threshold density display a similar density and energy distribution. We recover the same disk- and bar-like structures when using the same threshold density, independent of the method used to model the central object. The main difference can be seen in the bottom panel, where multiple sinks have been created. The gravitational interaction between them causes the surrounding gas to heat up, an effect that is not seen in the case with an adiabatic core. However, this difference does not significantly affect the evolution and accretion history of the central object.


\label{lastpage}
\end{document}